\documentclass[iop]{emulateapj}

\shorttitle{Elemental Assay of VMP, EMP, and UMP Stars}
\shortauthors{Hansen et al.}

\begin{document}

\title{An Elemental Assay of Very, Extremely, and Ultra Metal-Poor Stars\footnotemark[1]}
\footnotetext[1]{Based on observations made with the
    European Southern Observatory telescopes.}

\author{T. Hansen,\altaffilmark{1} C. J. Hansen,\altaffilmark{2}
  N. Christlieb,\altaffilmark{1} T. C. Beers,\altaffilmark{3}
  D. Yong,\altaffilmark{4} M. S. Bessell,\altaffilmark{4}
  A. Frebel,\altaffilmark{5} A. E. Garc\'{i}a P\'{e}rez,\altaffilmark{6}
  V. M. Placco,\altaffilmark{3} J. E. Norris,\altaffilmark{4} M. Asplund\altaffilmark{4}}

\altaffiltext{1}{Landessternwarte, ZAH, K{\"o}nigstuhl 12, 69117 Heidelberg,
  Germany, thansen@lsw.uni-heidelberg.de}
\altaffiltext{2}{Dark Cosmology Centre, Niels Bohr Institute, University of Copenhagen, Juliane Maries Vej 30, DK-2100 Copenhagen \O, Denmark}
\altaffiltext{3}{Department of Physics and JINA-CEE: Joint Institute for Nuclear Astrophysics -- Center 
for the Evolution of the Elements, University of Notre Dame, Notre Dame, IN 46556, USA}
\altaffiltext{4}{Research School of Astronomy and Astrophysics, The Australian National
  Univeristy, Weston, ACT 2611, Australia}
\altaffiltext{5}{Kavli Institute for Astrophysics and Space Research and Department of Physics, Massachusetts Institute of Technology, Cambridge, MA 02139, USA}
\altaffiltext{6}{Instituto de Astrof\'{\i}sica de Canarias, E38205 La 
Laguna, Tenerife, Spain and Departamento de Astrof\'{\i}sica, Universidad de La 
Laguna, 38206 La Laguna, Tenerife, Spain}

\begin{abstract}

We present a high-resolution elemental-abundance analysis for a sample
of 23 very metal-poor (VMP; $\mathrm{[Fe/H]} < -2.0$) stars, 12 of which
are extremely metal-poor (EMP; $\mathrm{[Fe/H]} < -3.0$), and 4 of which
are ultra metal-poor (UMP; $\mathrm{[Fe/H]} < -4.0$). These stars were
targeted to explore differences in the abundance ratios for elements
that constrain the possible astrophysical sites of element production,
including Li, C, N, O, the $\alpha$-elements, the iron-peak elements,
and a number of neutron-capture elements. This sample substantially
increases the number of known carbon-enhanced metal-poor (CEMP) and
nitrogen-enhanced metal-poor (NEMP) stars -- our program stars include
eight that are considered ``normal'' metal-poor stars, six CEMP-$no$
stars, five CEMP-$s$ stars, two CEMP-$r$ stars, and two CEMP-$r/s$
stars. One of the CEMP-$r$ stars and one of the CEMP-$r/s$ stars are
possible NEMP stars. We detect lithium for three of the six CEMP-$no$
stars, all of which are Li-depleted with respect to the Spite plateau.
The majority of the CEMP stars have $\mathrm{[C/N]}>0$. The stars
with $\mathrm{[C/N]}<0$ suggest a larger degree of mixing; the few
CEMP-$no$ stars that exhibit this signature are only found at
$\mathrm{[Fe/H]}<-3.4$, a metallicity below which we also find the
CEMP-$no$ stars with large enhancements in Na, Mg, and Al. We confirm
the existence of two plateaus in the absolute carbon abundances of CEMP
stars, as suggested by Spite et al. We also present evidence for a
``floor'' in the absolute Ba abundances of CEMP-$no$ stars at $A(\rm Ba)
\sim -2.0$.   

\end{abstract}


\keywords{Early universe: general --- Galaxy: formation --- Galaxy: halo --- nuclear reactions,
  nucleosynthesis, abundances --- stars: abundances }

\section{Introduction}

In recent years, high-resolution spectroscopic analyses of samples of
stars with metallicities significantly below solar have grown to the
point that one can begin to establish the general behaviors of elemental
abundance ratios associated with production by the first few generations
of stars to form the Galaxy (for a recent review see, e.g., Frebel \&
Norris 2015). These ``statistical'' samples are particularly valuable
when the data are analysed in a self-consistent manner
\citep[e.g.][]{yong2013}, so that comparisons of derived abundance
ratios are not plagued by the scatter introduced from the different
assumptions and procedures used by individual researchers, which can be
sufficiently large as to obscure important details.  

Of particular interest to this effort is the class of stars that,
despite their overall low abundances of iron-peak elements, exhibit
large over-abundances of C (as well as N and O) in their atmospheres,
the so-called carbon-enhanced metal-poor (CEMP) stars \citep{beers1992,
beerschristlieb2005, norris2013b}. This class comprises a number of
sub-classes (originally defined by Beers \& Christlieb 2005), based on
the behavior of their neutron-capture elements: (1) CEMP-$no$ stars,
which exhibit no over-abundances of n-capture elements, (2) CEMP-$s$
stars, which show n-capture over-abundances consistent with the slow
neutron-capture process, (3) CEMP-$r$ stars, with n-capture
over-abundances associated with the rapid neutron-capture process, and
(4) CEMP-$r/s$ stars, which exhibit n-capture over-abundances that
suggest contribution from both the slow and rapid neutron-capture
processes. Each of these sub-classes appear to be associated with
different element-production histories, thus their study provides
insight into the variety of astrophysical sites in the early Galaxy that
were primarily responsible for their origin. The CEMP-$no$ stars are of
special importance, as the preponderance of evidence points to their
being associated with elemental-abundance patterns that were produced by
the very first generation of massive stars \citep{norris2013b,
hansen2014, maeder2014}, thus they potentially provide a unique probe of
the first mass function in the early universe along with providing
information on the nucleosynthesis and properties of the first stars.

In a previous paper, \citet{hansen2014} (hereafter paper~I) provided a
detailed study of the elemental abundances for a sample of four ultra
metal-poor stars with $\mathrm{[Fe/H]} < -4.0$, three of which are clear
examples of CEMP-$no$ stars. Here we supplement this sample with an
additional 19 stars, exploring a wider range of metallicity. This allows
for the inclusion of additional examples of CEMP-$no$, CEMP-$s$,
CEMP-$r$, and CEMP-$r/s$ stars (two of which qualify as possible
nitrogen-enhanced metal-poor (NEMP) stars), providing a more complete
picture of the variety of elemental-abundance patterns for stars of very
low metallicity. 

This paper is outlined as follows. Section 2 summarizes our observations
and data analysis techniques. Section 3 presents our abundance analysis
results, while Section 4 provides a summary and brief discussion of
their implications.

\section{Observations and Data Analysis}

Our sample of 23 very metal-poor (VMP; [Fe/H] $\le -2.0$), extremely
metal-poor (EMP; [Fe/H] $\le -3.0$), ultra metal-poor (UMP; [Fe/H] $\le
-4.0$) stars presented here were originally selected from the
Hamburg/ESO Survey \citep[HES;][]{christlieb2008,frebel2006}, followed
up with medium-resolution spectroscopy on a variety of 2-m to 4-m class
telescopes (AAT~3.9m, CTIO~4m, CTIO~1.5m, ESO~3.6m, KPNO~4m,
SOAR~4m, SSO~2.3m, and UKST~1.2m), and then observed at high spectral
resolution with VLT/UVES \citep{dekker2000}. Paper~I describes the
observations and analysis of the four UMP stars in this sample.  

The high-resolution spectroscopy of the stars in our sample was
performed with UVES using the dichroic (DIC) beam splitter, allowing
simultaneous observation with the blue and red arm, in order to cover a
spectral range including a large number of chemical elements.
Three different settings were used: DIC (blue central wavelength + red
central wavelength), covering the following wavelengths -- DIC1
(390+580) blue: $\lambda$3260-4450\,{\AA}, red: $\lambda$4760-6840\,
{\AA}, DIC2 (346+760) blue: $\lambda$3030-3880\, {\AA}, red:
$\lambda$5650-9460\, {\AA}, and DIC2 (437+760) blue:
$\lambda$3730-4990\, {\AA}, red: $\lambda$5650-9460\,{\AA}. The spectral
resolving power varies with the choice of wavelength setting and slit
width. The average resolving power of the spectra is R $\sim$ 45,000.
Positions, observation dates, exposure times, and specific settings for
the individual stars in the sample are listed in Table~\ref{tab1}. 

The spectra were reduced using the UVES reduction pipeline version
4.9.8. Radial-velocity shifts of the spectra were obtained using the
IRAF\footnote{IRAF is distributed by the National Astronomy Observatory,
Inc., under cooperative agreement with the National Science Foundation.}
task FXCOR. Individual spectra were cross-correlated with a template
spectrum obtained during the same observation run. For the 2005 run,
 HE~0134$-$1519 and HD~2796 were used as templates, for which we
find $V_r$ = 244.0 km~s$^{-1}$ and $V_r = -14.7$ km~s$^{-1}$,
respectively. For the 2006 run, HD~140283 was used, for which we
find $V_r = -185.4$ km~s$^{-1}$. For stars with multiple observations,
the individual spectra were co-added with the IRAF SCOMBINE task.
Finally the radial-velocity shifted (and combined) spectrum was
normalized. Table~\ref{tab2} lists the derived radial velocities and
signal-to-noise (S/N) ratios at specific wavelengths for the different
spectra. When a wavelength region is covered by more than one setting,
the one having the highest S/N ratio is listed. Note that, because the
spectra were only obtained spanning at most a few nights, these data are
not suitable for evaluation of the binary nature of our stars. However,
the high accuracy of our derived radial velocities (typically better
than 1 km~s$^{-1}$) should prove useful for comparison with future
binarity studies.
 
Three of the stars in our sample are re-discoveries and have radial
velocities reported in the literature. These three stars are;
HE~0054$-$2542 (CS~22942$-$019, CD-26:304), HE~0411$-$3558
(CS~22186$-$005) and HE~0945$-$1435. \citet{prestonsneden2001} found
HE~0054$-$2542 to be in a binary system with a period of 2800 days,
while \cite{norris1996} reports $V_r = 192.4$ km~s~$^{-1}$ for
HE~0411$-$3558, close to our value of $V_r = 196.2$ km~s~$^{-1}$, and
\cite{norris2013a} reports $V_r = 121.8.4$ km~s~$^{-1}$ for
HE~0945$-$1435, where we find $V_r = 144.8$ km~s~$^{-1}$, suggesting
that it is a likely binary star.

\begin{deluxetable*}{lrrrrrr}
\tablecaption{Observation Log\label{tab1}}
\tablewidth{0pt}
\tablehead{
\colhead{Stellar ID} & \colhead{RA} & \colhead{Dec} & \colhead{Date} &
\colhead{Exp.(s)} & \colhead{Slit($\arcsec$)\tablenotemark{a}} & \colhead{Setting\tablenotemark{b}}} 
\startdata
\object{HE 0010$-$3422} & 00 13 08.9 & -34 05 55 & 2005 Nov 18 & 3600 &1.2/1.2& 390/580 \\
\object{HE 0054$-$2542}\tablenotemark{c,e} & 00 57 18.0 & -25 26 09 & 2005 Nov 19 & 1200 &1.0/1.0& 390/580 \\  
\object{HE 0100$-$1622} & 01 02 58.5 & -16 06 31 & 2005 Nov 18 & 3600 &1.2/1.2& 390/580 \\
\object{HE 0109$-$4510} & 01 12 08.1 & -44 54 16 & 2005 Nov 18 & 3600 &1.0/1.0& 390/580 \\
\object{HE 0134$-$1519} & 01 37 05.4 & -15 04 23 & 2005 Nov 17 & 3600 &1.0/1.0& 390/580 \\
                      &            &           & 2005 Nov 20 & 10800&0.8/0.8& 346/760 \\
\object{HE 0233$-$0343} & 02 36 29.7 & -03 30 06 & 2005 Nov 17 & 3600 &0.8/0.8& 390/580 \\ 
                      &            &           & 2005 Nov 18 & 7200 &1.0/0.7& 390/580 \\ 
                      &            &           & 2005 Nov 19 & 4940 &0.8/0.7& 346/760 \\ 
                      &            &           & 2005 Nov 20 & 3600 &0.8/0.7& 346/760 \\ 
\object{HE 0243$-$3044} & 02 45 16.4 & -30 32 02 & 2005 Nov 18 & 5400 &1.0/1.0& 390/580 \\
                      &            &           & 2005 Nov 20 & 3600 &0.8/0.7& 437/760 \\
\object{HE 0411$-$3558}\tablenotemark{d,e} & 04 13 09.0 & -35 50 39 & 2005 Nov 17 & 3600 &0.4/0.3& 390/580 \\
\object{HE 0440$-$1049}\tablenotemark{e} & 04 42 39.7 & -10 43 24 & 2006 Apr 18 & 900  &1.2/1.2& 390/580 \\
\object{HE 0440$-$3426}\tablenotemark{e} & 04 42 08.1 & -34 21 13 & 2006 Apr 17 & 900  &1.2/1.2& 390/580 \\
\object{HE 0448$-$4806}\tablenotemark{e} & 04 49 33.0 & -48 01 08 & 2006 Apr 18 & 840  &1.2/1.2& 390/580 \\
\object{HE 0450$-$4902} & 04 51 43.3 & -48 57 25 & 2006 Apr 17 & 3600 &1.2/1.2& 390/580 \\
\object{HE 0945$-$1435}\tablenotemark{f} & 09 47 50.7 & -14 49 07 & 2006 Apr 17 & 3600 &1.2/1.2& 390/580 \\
                      &            &           & 2006 Apr 18 & 7200 &0.8/0.7& 437/760 \\
\object{HE 1029$-$0546} & 10 31 48.2 & -06 01 44 & 2006 Apr 17 & 3950 &1.2/1.2& 390/580 \\ 
                      &            &           & 2006 Apr 18 & 7200 &0.8/0.7& 390/580 \\ 
\object{HE 1218$-$1828} & 12 21 19.3 & -18 45 34 & 2006 Apr 21 & 6000 &1.2/1.2& 390/580 \\
                      &            &           & 2006 Apr 21 & 3000 &0.8/0.7& 390/580 \\
\object{HE 1241$-$2907} & 12 44 13.0 & -29 23 47 & 2006 Apr 17 & 3600 &1.2/1.2& 390/580 \\
\object{HE 1310$-$0536} & 13 13 31.2 & -05 52 13 & 2006 Apr 17 & 3600 &1.2/1.2& 390/580 \\
                      &            &           & 2006 Apr 18 & 7200 &0.8/0.7& 437/760 \\
\object{HE 1429$-$0347}\tablenotemark{e} & 14 32 26.1 & -04 00 31 & 2006 Apr 17 & 1800 &1.2/1.2& 390/580 \\
                      &            &           & 2006 Apr 18 & 7200 &0.8/0.7& 437/760 \\ 
\object{HE 2159$-$0551}\tablenotemark{e} & 22 02 16.4 & -05 36 48 & 2006 Apr 18 & 900  &0.8/0.7& 390/580 \\
\object{HE 2208$-$1239}\tablenotemark{e} & 22 10 53.3 & -12 24 27 & 2006 Apr 18 & 600  &0.8/0.7& 390/580 \\
\object{HE 2238$-$4131} & 22 41 22.6 & -41 15 57 & 2005 Nov 19 & 3600 &1.2/1.2& 390/580 \\
\object{HE 2239$-$5019} & 22 42 26.9 & -50 04 01 & 2005 Nov 17 & 10800&1.0/0.8& 390/580 \\
\object{HE 2331$-$7155} & 23 34 36.1 & -71 38 51 & 2005 Nov 17 & 3600 &0.8/0.8& 390/580 \\  
                      &            &           & 2005 Nov 20 & 7200 &0.8/0.7& 437/760 \\   
\enddata
\tablenotetext{a}{Blue slit / red slit.}
\tablenotetext{b}{Spectrograph setting 390/580 = DIC1 (390+580), etc.}
\tablenotetext{c}{CS 22942-019; CD-26:304.}
\tablenotetext{d}{CS 22186-005.}
\tablenotetext{e}{\citet{frebel2006}.}
\tablenotetext{f}{\citet{norris2013a}.}
\end{deluxetable*}

\begin{deluxetable*}{lrrrrr}
\tablecaption{Radial Velocities and Signal-To-Noise Ratios\label{tab2}}
\tablewidth{0pt}
\tablehead{
\colhead{} & \colhead{$V_r$} & \colhead{$V_{r,\rm err}$} & \colhead{S/N} & \colhead{S/N} & \colhead{S/N}\\
\colhead{Stellar ID} & \colhead{(km~s$^{-1}$)} & \colhead{(kms~$^{-1}$)} &
\colhead{$\lambda3400$\,{\AA}} & \colhead{$\lambda4300$\,{\AA}} &
\colhead{$\lambda6700$\,{\AA}}} 
\startdata
\object{HE 0010$-$3422} & 158.8   & 0.2 & 11 & 49  & 84  \\
\object{HE 0054$-$2542} &$-$214.6 & 0.1 & 9  & 46  & 96  \\  
\object{HE 0100$-$1622} &  28.6   & 0.3 & 3  & 17  & 39  \\
\object{HE 0109$-$4510} & 138.8   & 0.1 & 5  & 25  & 33  \\
\object{HE 0134$-$1519} & 244.0   & 1.0 & 14 & 54  & 75  \\
\object{HE 0233$-$0343} &  63.5   & 0.6 & 9  & 35  & 42 \\ 
\object{HE 0243$-$3044} &  39.8   & 0.3 & 9  & 14  & 32  \\
\object{HE 0411$-$3558} & 196.2   & 0.3 & 26 & 105 & 110 \\
\object{HE 0440$-$1049} & 158.9   & 3.0 & 16 & 65  & 86 \\
\object{HE 0440$-$3426} & 326.0   & 0.6 & 16 & 61  & 162 \\
\object{HE 0448$-$4806} & 133.5   & 0.7 & 10 & 44  & 68 \\
\object{HE 0450$-$4902} & 332.4   & 1.5 & 4  & 26  & 29 \\
\object{HE 0945$-$1435} & 144.8   & 0.4 & 12 & 44  & 80  \\
\object{HE 1029$-$0546} &  18.6   & 0.3 & 10 & 35  & 45\\ 
\object{HE 1218$-$1828} & 147.1   & 0.5 & 4  & 19  & 33  \\
\object{HE 1241$-$2907} & 336.3   & 2.2 & 4  & 31  & 16  \\
\object{HE 1310$-$0536} & 113.2   & 1.7 & 1  & 39  & 65  \\
\object{HE 1429$-$0347} &$-$143.3 & 0.4 & 3  & 71  & 129 \\ 
\object{HE 2159$-$0551} &$-$131.3 & 0.8 & 2  & 50  & 72  \\
\object{HE 2208$-$1239} & $-$43.1 & 0.6 & 5  & 55  & 102 \\
\object{HE 2238$-$4131} & $-$42.0 & 0.3 & 2  & 13  & 32  \\
\object{HE 2239$-$5019} & 368.7   & 0.5 & 9  & 44  & 43 \\
\object{HE 2331$-$7155} & 210.6   & 0.8 & 6  & 51  & 120   
\enddata
\end{deluxetable*}

\subsection{Stellar Parameters}

The stellar atmospheric parameters were determined following most of the
steps outlined in \citet{yong2013} and in Paper~I, so that the results
of the abundance analyses of their sample and ours can be usefully
combined. 

The effective temperature ($T_{\rm eff}$) was, for the majority of the
stars, determined by fitting spectrophotometric observations of the star
with model-atmosphere fluxes \citep{bessell2007,norris2013a}. For this
step, medium-resolution spectra were obtained with the Wide Field
Spectrograph \citep[WiFeS; ][]{dopita2007} on the Australian National
University 2.3-m telescope at Siding Spring Observatory during 2012.
This is a double-beam spectrograph using a dichroic mirror to separate
the blue and red regions. The spectrograph covers the wavelength ranges
3000-6200\,{\AA} and 6000-9700\,{\AA} in the blue and red, respectively,
with a resolution of 2\,{\AA}. The observations, data reduction, and
analysis were performed as described in Section 4.1 of
\citet{norris2013a}. The reduced spectra were cross-correlated against a
grid of MARCS atmosphere models \citep{gustafsson2008} using the PHYTON
program fitter, written by S.J. Murphy. The MARCS models have parameters
ranging in $T_{\rm eff}$ from 2500~K to 8000~K, in steps of 100~K from
2500~K to 4000~K, and in steps of 250~K from 4000~K to 8000~K. Surface
gravity ($\log g$) values for the grid were between $-1.0$ (cgs) and 5.5
(cgs) in steps of 0.5 dex, and metallicities between $-5.0$ and $+1.0$
in variable steps. As the stars in this sample all have very low
metallicities, $\alpha$-enhanced models were used, with
$\mathrm{[\alpha/Fe]}=+0.25$ for $-1.5\leq\mathrm{[Fe/H]}\leq-0.5$ and
$\mathrm{[\alpha/Fe]}=+0.4$ for $-5.0\leq\mathrm{[Fe/H]}<-1.5$.

For two stars in the sample (marked in Table~\ref{tab3}) we did not have
spectrophotometric observations. The effective temperatures for these
stars were determined from broadband photometry, using the $V-K$ color
index, as this is least affected by metallicity \citep{alonso1999}. The
$V$ and $K$ magnitudes for the stars are listed in Table~\ref{tab3}. The
2MASS $K$ magnitudes were converted to the Johnson photometric system
using the filter conversion $K_{\rm Johnson}$ = $K_{\rm 2MASS}+0.044$
\citep{bessell2005}. Reddening values, $E(B-V)$, are adopted from
\citet{schlegel1998}; values exceeding 0.1 mag were corrected
according to \citet{bonifacio2000}. These values were then converted to
$E(V-K)$, using the relation from \citet{alonso1996}, $E(V-K)=2.72\,
E(B-V)$. The final de-reddened $V-K$ colors were thus found from the
following equation: $V-K_{0,\rm Johnson}=V_{\rm Johnson}-K_{\rm
2MASS}+0.044-2.72\,E(B-V)$. 

To estimate the effective temperatures we used the calibration of
\citet{alonso1996}, as this provides temperatures that are in good agreement with the
scale used for the majority of our sample. We determined temperatures
using this method for as many stars in the sample as possible, in order
to estimate the offset between the two temperature scales. We found an
average offset of $+30$~K between the two temperature scales, and have
corrected the temperatures determined from the $V-K$ colors accordingly
($T_{\rm eff} $=$ T_{{\rm eff},V-K} + 30$). The $V$ and $B-V$ photometry
listed for HE~0010$-$3422 in Table~\ref{tab3} is almost certainly in
error, as it results in a temperature $\sim 1500$~K below what was found
from the spectrophotometric observations. Due to this large difference,
this star has been excluded from the determination of the offset between
the two temperature scales. 

Surface gravity ($\log g$) estimates for the stars were determined from
the $Y^2$ isochrones \citep{demarque2004}, assuming an age of 10 Gyr
\citep{yong2013} and an $\alpha$-element enhancement of
$\mathrm{[\alpha/Fe]}=+0.3$ (the isochrones exists with
$\mathrm{[\alpha/Fe]}=0.0$, $\mathrm{[\alpha/Fe]}=+0.3$ and
$\mathrm{[\alpha/Fe]}=+0.6$). The isochrones extend in metallicity down
to $\mathrm{[Fe/H]}=-3.5$, so for the six stars in the sample with
metallicities in the range $-4.7\leq\mathrm{[Fe/H]}\leq-3.5$, a linear
extrapolation down to $\mathrm{[Fe/H]}=-4.7$ has been used to obtain the
gravity estimate. The average difference between the listed surface
gravities where the actual $\mathrm{[Fe/H]}$ have been used and the
surface gravity obtained using the $\mathrm{[Fe/H]}=-3.5$ isochrone is
small, on the order of 0.05 dex. 

\begin{figure}
\includegraphics[angle=0,width=3.5in]{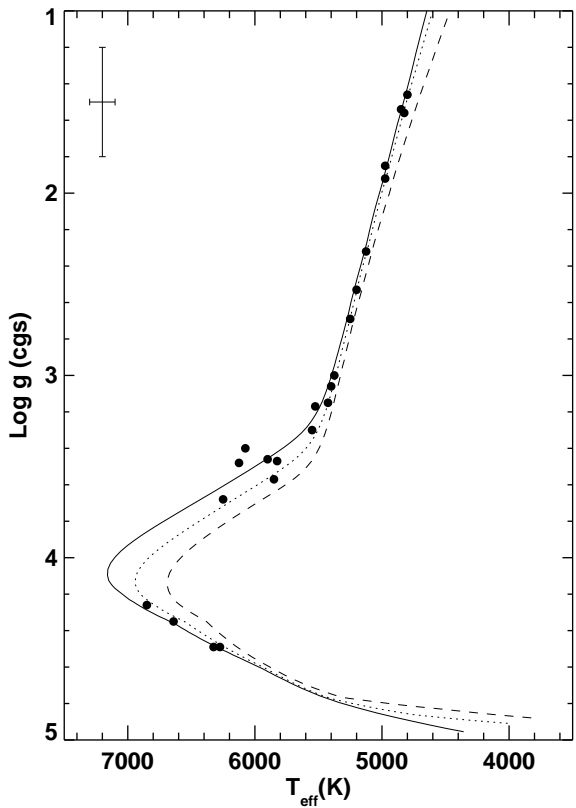}
\caption{$T_{\rm eff}$ vs. $\log g$ diagram for the program stars, over-plotted
with 10 Gyr isochrones for three different metallicities: $\mathrm{[Fe/H]}=-3.5$
(solid line), $\mathrm{[Fe/H]}=-2.5$ (dotted line) and $\mathrm{[Fe/H]}=-1.9$
(dashed line). All of the isochrones have $\mathrm{[\alpha/Fe]}= +0.3$. A
representative error bar on the derived parameters is shown in the upper left. \label{fig1}}
\end{figure}

Figure~\ref{fig1} shows the $T_{\rm eff}$ vs. $\log g$ diagram for the
program stars with isochrones for three different metallicities:
$\mathrm{[Fe/H]}=-1.9$, $\mathrm{[Fe/H]}=-2.5$, and
$\mathrm{[Fe/H]}=-3.5$. All isochrones have $\mathrm{[\alpha/Fe]}= +0.3$
and an age of 10 Gyr. The sample shows a mixture of dwarfs,
subgiants, and giants.

For five of the warmer stars in the sample, HE~0233$-$0343,
HE~0411$-$3558, HE~0450$-$4209, HE~1241$-$2907, and HE~2239$-$5019, the
isochrones returned two possible solutions for the gravity. For these
five stars we have tried to derive spectroscopic gravities and/or
checked the isochrone gravities by fitting profiles of gravity-sensitive
lines. The gravity of stars can be determined spectroscopically by
enforcing ionization equilibrium between lines formed by neutral atoms
and lines formed by ions, e.g., Fe~I and Fe~II or Ti~I and Ti~II, taking
advantage of the fact that Fe~I and Ti~I lines are not significantly
gravity sensitive, while the Fe~II and Ti~II lines are. We
performed this analysis for the stars where both Fe~I and Fe~II or Ti~I
and Ti~II lines were detected. A check of the gravity can also be
performed by fitting the profiles of gravity-sensitive lines. Lines such
as Mg~I and the Ca~II H and K lines exhibit strong pressure-broadened
wings in cool stars. We performed spectral syntheses of these lines
using model atmospheres with the two possible gravities, keeping all
other parameters constant, in order to see which of the two possible
gravities yields the best fit in the wings of these lines. The result of
these tests yielded subgiant gravities for HE~0233$-$0343,
HE~0411$-$3558 and HE~2239$-$5019 and dwarf gravities for HE~0450$-$4209
and HE~1241$-$2907 (see Table~\ref{tab3}).

The microturbulent velocity ($\xi$) was computed in the usual way, by forcing
the abundances from individual Fe~I lines to show no trend with reduced
equivalent width, $\log(W_{\lambda}/\lambda)$. For HE~0233$-$0343, too few
Fe~I line were present to determine the microturbulent velocity in this
manner, so a fixed valued of $\xi = 2$~km s$^{-1}$ was used for this
star, following paper~I.

Metallicities were determined from equivalent-width measurements of the
Fe~I lines. For a few stars we also detected a number of Fe~II lines;
for these stars there is good agreement between the abundance derived
from the Fe~II lines and that from the Fe~I lines used for determining
the temperature, gravity, and microturbulence. 

The final stellar parameters and their estimated uncertainties are listed
in Table~\ref{tab3}.

\begin{deluxetable*}{lcccccccccc}
\tablecaption{Stellar Photometry and Stellar Atmospheric Parameters \label{tab3}}
\tablewidth{0pt}
\tablehead{
\colhead{Star} & \colhead{$V$} & \colhead{$B-V$} &
\colhead{R\tablenotemark{a}} &
\colhead{$E(B-V)$} & \colhead{$K$\tablenotemark{b}} & \colhead{$(V-K)_0$} &
\colhead{$T_{\rm eff}$} & \colhead{$\log g$} & \colhead{$\mathrm{[Fe/H]}$} & \colhead{$\xi$}\\
\colhead{} & \colhead{} & \colhead{} & \colhead{} & \colhead{} & \colhead{} &
\colhead{} &
\colhead{($\pm$100~K)} & \colhead{($\pm$0.3~dex)} &
\colhead{($\pm$0.2~dex)} & \colhead{($\pm$0.3~km s$^{-1}$)}}
\startdata
\object{HE 0010$-$3422\tablenotemark{c}} & 15.48 & 0.095 & 1 & 0.017 & 12.34 &\nodata& 5400 & 3.1 & $-$2.7 & 2.4\\
\object{HE 0054$-$2542} & 12.69 & 0.880 & 2 & 0.040 & 10.65 & 1.89  & 5300 & 2.7 & $-$2.5 & 1.3\\ 
\object{HE 0100$-$1622} & 15.82 & 0.837 & 1 & 0.021 & 13.85 & 1.87  & 5400 & 3.0 & $-$2.9 & 1.1\\
\object{HE 0109$-$4510} & 16.01 & 0.523 & 1 & 0.011 & 14.18 & 1.75  & 5600 & 3.3 & $-$3.0 & 1.1\\
\object{HE 0134$-$1519} & 14.47 & 0.463 & 1 & 0.016 & 12.68 & 1.71  & 5500 & 3.2 & $-$4.0 & 1.5\\
\object{HE 0233$-$0343} & 15.43 & 0.437 & 1 & 0.025 & 14.06 & 1.26  & 6100 & 3.4 & $-$4.7 & 2.0\\
\object{HE 0243$-$3044} & 16.13 & 0.833 & 1 & 0.019 & 14.17 & 1.83  & 5400 & 3.2 & $-$2.6 & 0.9\\
\object{HE 0411$-$3558} & 12.96 & 0.382 & 1 & 0.011 & 11.58 & 1.31  & 6300 & 3.7 & $-$2.8 & 3.4\\
\object{HE 0440$-$1049} &\nodata&\nodata&   & 0.107 &\nodata&\nodata& 5800 & 3.5 & $-$3.0 & 1.6\\
\object{HE 0440$-$3426} & 11.44 & 1.440 & 3 & 0.013 &  8.97 & 2.39  & 4800 & 1.6 & $-$2.2 & 1.9\\
\object{HE 0448$-$4806} & 12.35 & 1.100 & 3 & 0.021 & 11.21 & 1.04  & 5900 & 3.6 & $-$2.3 & 1.2\\
\object{HE 0450$-$4902} &\nodata&\nodata&   & 0.009 &\nodata&\nodata& 6300 & 4.5 & $-$3.1 & 1.2\\ 
\object{HE 0945$-$1435} &\nodata&\nodata&   & 0.054 &\nodata&\nodata& 6300 & 4.5 & $-$3.9 & 1.6\\
\object{HE 1029$-$0546} & 15.63 & 0.355 & 1 & 0.043 & 14.37 & 1.10  & 6650\tablenotemark{d}& 4.3 & $-$3.3 & 1.6\\
\object{HE 1218$-$1828} & 16.34 & 0.493 & 1 & 0.043 & 14.70 & 1.48  & 5900\tablenotemark{d}& 3.5 & $-$3.4 & 1.8\\
\object{HE 1241$-$2907} &\nodata&\nodata&   & 0.071 &\nodata&\nodata& 6900 & 3.8 & $-$3.0 & 1.8\\
\object{HE 1310$-$0536} & 14.35 & 0.708 & 1 & 0.043 & 11.90 & 2.29  & 5000 & 1.9 & $-$4.2 & 2.2\\
\object{HE 1429$-$0347} & 13.69 & 0.687 & 1 & 0.110 & 11.41 & 1.94  & 5000 & 1.9 & $-$2.7 & 1.5\\
\object{HE 2159$-$0551} &\nodata&\nodata&   & 0.060 &  9.85 &\nodata& 4800 & 1.5 & $-$2.8 & 2.1\\
\object{HE 2208$-$1239} &\nodata&\nodata&   & 0.041 &\nodata&\nodata& 5100 & 2.3 & $-$2.9 & 2.0\\
\object{HE 2238$-$4131} & 16.10 &       &   & 0.013 & 13.85 & 2.17  & 5200 & 2.5 & $-$2.8 & 1.0\\
\object{HE 2239$-$5019} & 15.85 & 0.393 & 1 & 0.009 & 14.28 & 1.50  & 6100 & 3.5 & $-$4.2 & 1.8\\
\object{HE 2331$-$7155} & 11.73 &       &   & 0.032 &\nodata&\nodata& 4900 & 1.5 & $-$3.7 & 2.2\\
\enddata
\tablenotetext{a}{Source of $V$ and $B-V$ color; 1 = \citet{beers2007}, 
2 = \citet{rossi2005}, 3 = \citet{hoeg2000}.}
\tablenotetext{b}{2MASS Catalog \citep{skrutskie2006}.}
\tablenotetext{c}{The $V$ and $B-V$ for this star reported by
\citet{beers2007} is almost certainly in error.}
\tablenotetext{d}{Photometric temperature.}
\end{deluxetable*}

\subsection{Abundances}

The elemental abundances were derived by synthesizing individual
spectral lines and molecular bands. All abundances are derived under the
assumption of 1D and Local Thermodynamic Equilibrium, and adopting the
solar abundances from \citet{asplund2009}. 

The 2011 version of MOOG \citep{sobeck2011,sneden1973} was used for the
synthesis; this version of MOOG includes proper treatment of continuum
scattering. For stars in the temperature range of our sample, the two
main sources of opacity in stellar atmospheres are bound-free absorption
from the negative hydrogen ion (H$^-$) and Rayleigh scattering from
neutral atomic hydrogen. Their individual contributions to the total
opacity depends on temperature and metallicity; at low temperature and
low metallicity the contribution from Rayleigh scattering is almost
equal to the contribution from bound-free absorption. So, when working
with metal-poor stars it is especially important to model the scattering
accurately to obtain the correct line intensities. 

To perform the synthesis we used the $\alpha$-enhanced NEWODF grid of
ATLAS9 atmosphere models \citep{castelli2003}, interpolated with
software developed by C. Allende Prieto, to obtain the models matching
the parameters of the stars \citep[e.g.,][]{reddy2003,allende2004}. The
$\alpha$-enhanced ATLAS9 models cover a range in $T_{\rm eff}$, from
3500~K to 50000~K, $\log g$, from 0.0 to 5.0 (cgs), for metallicities,
$\mathrm{[Fe/H]}$, in the range $-2.5$ to $+$0.5 and
$\mathrm{[Fe/H]=-4.0}$. For the metallicity $\mathrm{[Fe/H]} = -3.5$,
models exists with temperatures in the range 3500~K to 6500~K, and
surface gravities ranging from 0.0 to 5.0.

We used atomic data from the Gaia/ESO line list version 4 (Heiter et
al., in prep.) for the analysis. This list covers the lines between
4750\,{\AA} to 6850\,{\AA} and 8500\,{\AA} to 8950\,{\AA}. For lines not
covered by the Gaia/ESO line list, atomic data from the VALD database
\citep{kupka2000} were adopted. A number of the elements analyzed
exhibit hyperfine splitting (Sc, Mn, Co, Y, Zr, La, Pr, Nd and Eu). For
those elements which only have lines in the region not covered by the
Gaia/ESO line list, hyperfine splitting from \cite{kurucz1995} was used.
The lines of Li, Ba, and Pb have both hyperfine splitting and isotopic
shifts. The lines of Li and Ba are included in the Gaia/ESO line list,
while Pb is not, so for this element data from \citet{simons1989} was
used. 

Carbon and nitrogen abundances (or upper limits), and the isotopic
ratios $^{12}$C/$^{13}$C, were obtained by synthesizing molecular bands,
namely the 4300\,{\AA} CH G-band, the NH band at 3360\,{\AA} and the CN
bands at 3890\,{\AA} and 4215\,{\AA}. All molecular information is taken
from \cite{masseron2014} and T. Masseron (priv. comm.). Dissociation
energies of 3.47~eV, 3.42~eV, and 7.74~eV were used for the species CH,
NH and CN, respectively.

\begin{figure}
\includegraphics[angle=0,width=3.5in]{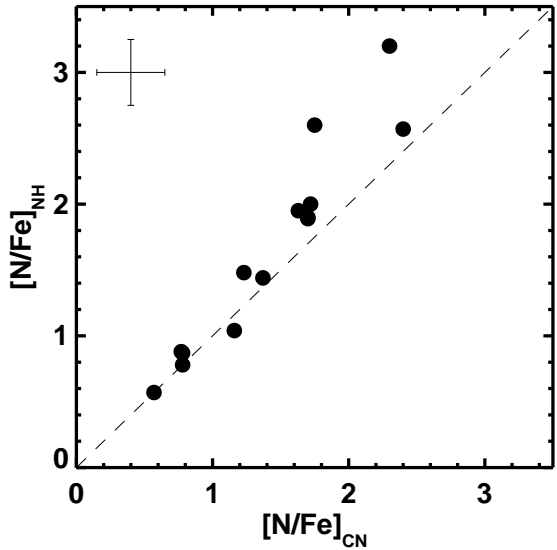}
\caption{Comparison of nitrogen abundances estimated from the NH (3360\,
{\AA}) and the CN (3890\,{\AA}) bands for stars where abundances could be 
derived from both bands. The dashed line is the one-to-one correlation. A
representative error bar on the abundances ratios is shown in the upper left. \label{figN}}
\end{figure}

When possible, N abundances were determined from synthesis of the CN
bands using the C abundances computed from the CH band as fixed input.
If no CN band was visible, the nitrogen abundance was derived from the
NH band (which falls in a region of the spectra with substantially lower
S/N). For stars where abundances could be derived from both bands the
resultant N abundances, derived from the NH and CN bands respectively,
are compared in Figure~\ref{figN}. For low N abundances, the results
from the two bands agree well, while a discrepancy is seen at high N,
with higher abundances being derived from the NH band compared to the
CN band. The reason for this discrepancy is not clear, but the physical
parameters, such as line positions and gf values, are less 
well-established for the NH band compared to the CN band, which could
account, at least in part, for it.

The carbon abundance is coupled to the oxygen abundance through the CO
molecule. Oxygen abundances or upper limits for the sample stars
are derived from the 6300\,{\AA} line, but for the majority of our
stars no reliable oxygen abundance could be obtained, so when deriving
the carbon abundances we used a typical halo-star value of
$\mathrm{[O/Fe]}= +0.4$ for oxygen.

\begin{figure}
\includegraphics[angle=0,width=3.5in]{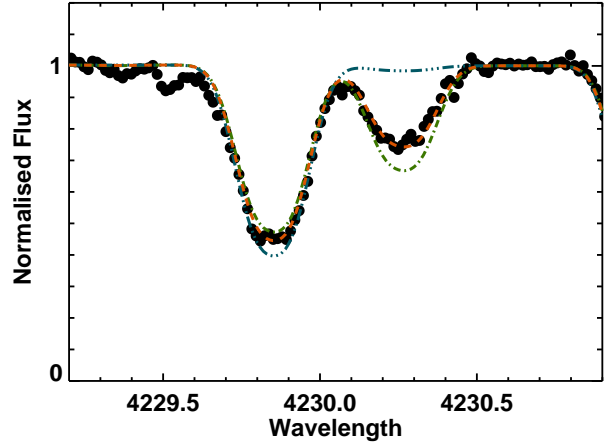}
\caption{Synthesis of the $^{13}$CH line at 4230\,{\AA} in HE~2208$-$1239 for
  three different isotopic ratios; 100\% $^{12}$CH (blue), $^{12}$C/$^{13}$C =
  79/21 (red) and $^{12}$C/$^{13}$C = 70/30 (green). \label{fig2}}
\end{figure}

Molecular $^{13}$CH features were identified for 11 of the stars in our
sample. The $^{12}$C/$^{13}$C isotopic ratios were determined from
the analysis of $^{13}$CH features in the wavelength range from 4210\,
{\AA} to 4235\,{\AA}. Figure~\ref{fig2} shows the synthesis of the
$^{13}$CH line at 4230\, {\AA} in HE~2208$-$1239 for three different
isotopic ratios.

The derived elemental abundances, along with propagated uncertainties
arising from the effects of uncertainties of the stellar parameters,
continuum placement, and line information, are listed in
Table~\ref{tab4} for the ``normal'' metal-poor (hereafter, NMP) stars,
Table~\ref{tab5} for the CEMP-$no$ stars, Table~\ref{tab6} for the
CEMP-$s$ stars, and Table~\ref{tab7} for the CEMP-$r$ and CEMP-$r/s$
stars. The $^{12}$C/$^{13}$C isotopic ratios, where available, are
listed in Table~\ref{tab8}.

\section{Results}

The abundance analysis yielded abundances or upper limits for 18
elements: Li, C, N, Na, Mg, Al, Si, Ca, Sc, Ti, Cr, Mn, Fe, Co, Ni, Sr,
Ba and Eu, for all the stars in the sample, plus abundances or
upper-limits of Y, Zr, La. Ce, Pr, Nd, Sm, Gd, Dy, Er and Pb, for the
CEMP-$s$, CEMP-$r$ and CEMP-$r/s$ stars in the sample. 

\begin{turnpage}
\begin{deluxetable*}{lrrrrrrrr}
\tablecaption{ Derived Abundances for ``Normal'' Metal-Poor (NMP) Stars\label{tab4}}
\tablewidth{0pt}
\tablehead{
\colhead{$\mathrm{[X/Fe]}$} & \colhead{\object{HE 0109$-$4510}}
&\colhead{\object{HE 0411$-$3558}} & \colhead{\object{HE 0945$-$1435}} &
\colhead{\object{HE 1218$-$1828}} & \colhead{\object{HE 1241$-$2907}} &
\colhead{\object{HE 1429$-$0347}} & \colhead{\object{HE 2159$-$0551}} &
\colhead{\object{HE 2239$-$5019}}} 
\startdata
$\mathrm{[Fe/H]} $& $-$2.96 (0.20)& $-$2.81 (0.20)& $-$3.87 (0.20)& $-$3.43
(0.20)& $-$3.00 (0.20)& $-$2.71 (0.20)& $-$2.81 (0.20)& $-$4.15 (0.20)\\
                  &        &        &        &        &        &       &       &       \\
A(Li)             & $+$1.46 (0.18)&$<+$1.44 \nodata& $+$1.88 (0.19)& $+$1.94
(0.17)&$<+$2.41 \nodata& $+$0.62 (0.16)&$<+$0.44 \nodata&$<+$1.70 \nodata\\
                  &        &        &        &        &        &       &       &       \\
$\mathrm{[C/Fe]} $& $+$0.43 (0.26) &$<+$0.70 \nodata&$<+$2.00 \nodata&$<+$1.27
\nodata&$<+$2.00 \nodata& $+$0.31 (0.23)& $-$0.24 (0.28)&$<+$1.70 \nodata\\
$\mathrm{[N/Fe]} $& $+$0.57 (0.25) &$<+$1.30 \nodata&$<+$2.10 \nodata&$<+$1.70
\nodata&$<+$2.60 \nodata& $+$1.89 (0.25)& $+$0.88 (0.29)&$<+$2.70 \nodata\\
$\mathrm{[O/Fe]} $&$<+$2.30 \nodata&$<+$2.30 \nodata&$<+$4.00 \nodata&$<+$3.30
\nodata&$<+$3.90 \nodata& $+$0.90 (0.15)&$<+$1.10 \nodata&$<+$4.00 \nodata\\
$\mathrm{[Na/Fe]}$& $-$0.10 (0.19) & $+$0.02 (0.15) &$<-$0.30 \nodata& $-$0.33
(0.16)& $+$0.87 (0.18)& $+$1.00 (0.32)& $+$0.14 (0.30)&$<-$0.30 \nodata\\
$\mathrm{[Mg/Fe]}$& $+$0.29 (0.19) & $+$0.24 (0.15) & $+$0.11 (0.14) & $+$0.32
(0.17)& $+$0.18 (0.19)& $+$0.28 (0.19)& $+$0.41 (0.24)& $+$0.45 (0.15)\\
$\mathrm{[Al/Fe]}$& $-$0.91 (0.20) & $-$0.59 (0.20) &$<-$0.50 \nodata& $-$0.57
(0.22)&$<-$0.60 \nodata& $+$0.00 (0.17)& $-$0.82 (0.23)& $-$0.57 (0.21)\\ 
$\mathrm{[Si/Fe]}$& $+$0.74 (0.20) & \nodata \nodata& $+$0.31 (0.16) & $+$0.06
(0.18)&$+$0.03 (0.20)& $+$0.66 (0.16)& $+$0.23 (0.22)& $+$0.06 (0.15)\\
$\mathrm{[Ca/Fe]}$& $+$0.26 (0.18) & $+$0.22 (0.15) & $+$0.12 (0.18) & $+$0.60
(0.17)&$+$0.21 (0.19)& $+$0.49 (0.18)& $+$0.17 (0.23)& $+$0.23 (0.15)\\
$\mathrm{[Sc/Fe]}$& $+$0.07 (0.22) & $+$0.38 (0.18) &$<-$0.10 \nodata& $+$0.12
(0.20)& $+$0.34 (0.22)& $-$0.08 (0.22)& $-$0.13 (0.27)& $+$0.32 (0.16)\\
$\mathrm{[Ti/Fe]}$& $+$0.22 (0.23) & $+$0.59 (0.14) & $-$0.04 (0.18) & $+$0.25
(0.16)& $+$0.28 (0.18)& $+$0.30 (0.18)& $+$0.24 (0.23)& $+$0.37 (0.17)\\
$\mathrm{[V/Fe]} $&$<+$0.80 \nodata&$<+$1.00 \nodata&$<+$1.80
\nodata&$<+$1.30\nodata&$<+$1.60 \nodata& $+$0.21 (0.17)& $+$0.17
(0.23)&$<+$2.00 \nodata\\
$\mathrm{[Cr/Fe]}$& $-$0.47 (0.18) & $-$0.05 (0.17) & $-$0.17 (0.16) & $-$0.28
(0.19)& $+$0.04 (0.21)& $-$0.36 (0.18)& $-$0.40 (0.24)& $+$0.00 (0.17)\\
$\mathrm{[Mn/Fe]}$& $-$1.02 (0.20) & $-$0.62 (0.20) &$<-$0.20 \nodata& $-$0.65
(0.22)&$<+$0.00 \nodata & $-$0.70 (0.20)& $-$0.45 (0.25)&$<-$0.10 \nodata\\
$\mathrm{[Co/Fe]}$& $+$0.47 (0.21) & $+$0.16 (0.19) &$<+$1.00 \nodata& $+$0.38
(0.21)&$<+$1.00 \nodata& $-$0.11 (0.19)& $-$0.06 (0.25)&$<+$1.00 \nodata\\
$\mathrm{[Ni/Fe]}$& $+$0.68 (0.21) & $+$0.10 (0.17) & $+$0.17 (0.14) & $-$0.06
(0.19)&$+$0.16 (0.21)& $+$0.00 (0.27)& $-$0.03 (0.30)& $+$0.24 (0.17)\\
$\mathrm{[Zn/Fe]}$&$<+$0.80 \nodata&$<+$0.60 \nodata&$<+$1.80 \nodata&$<+$1.50
\nodata&$<+$1.40 \nodata& $+$0.19 (0.15)& $+$0.36 (0.21)&$<+$2.00 \nodata\\
$\mathrm{[Sr/Fe]}$& $+$1.00 (0.21) & $-$0.54 (0.19) &$<-$0.90 \nodata& $-$0.51
(0.20)& $-$0.77 (0.22)& $+$0.10 (0.16)& $+$0.16 (0.22)&$<-$0.60 \nodata\\ 
$\mathrm{[Ba/Fe]}$&$<-$1.30 \nodata& $-$0.95 (0.14) & $<$0.00 \nodata&$<-$0.52
\nodata&$<-$0.30 \nodata& $-$0.34 (0.17)& $-$1.58 (0.23)&$<+$0.00 \nodata\\
$\mathrm{[Eu/Fe]}$&$<+$1.10 \nodata&$<+$1.50 \nodata&$<+$2.20 \nodata&$<+$2.00
\nodata&$<+$2.40 \nodata& $+$0.47 (0.20)& $-$0.20 (0.25)&$<+$2.50 \nodata\\
\enddata
\end{deluxetable*}
\end{turnpage}

\begin{deluxetable*}{lrrrrrr}
\tablecaption{Derived Abundances for CEMP-$no$ Stars\label{tab5}}
\tablewidth{0pt}
\tablehead{
\colhead{} & \colhead{\object{HE 0100$-$1622}} & \colhead{\object{HE 0134$-$1519}}
& \colhead{\object{HE 0233$-$0343}}
& \colhead{\object{HE 0440$-$1049}} & \colhead{\object{HE 1310$-$0536}} &
\colhead{\object{HE 2331$-$7155}}} 
\startdata
$\mathrm{[Fe/H]} $& $-$2.93 (0.20) & $-$3.98 (0.20) & $-$4.68 (0.20) & $-$3.02
(0.20)& $-$4.15 (0.20)& $-$3.68 (0.20)\\
                  &        &        &        &        &        &         \\
A(Li)             &$<+$1.12 \nodata& $+$1.27 (0.19) & $+$1.77 (0.18) & $+$2.00
(0.14)&$<+$0.80 \nodata&$<+$0.37 \nodata\\
                  &        &        &        &        &        &         \\
$\mathrm{[C/Fe]} $& $+$2.75 (0.29) & $+$1.00 (0.26) & $+$3.48 (0.24) & $+$0.69
(0.25)& $+$2.36 (0.23)& $+$1.34(0.26)\\
$\mathrm{[N/Fe]} $& $+$1.90 (0.28) &$<+$1.00 \nodata&$<+$2.80 \nodata&$<+$0.62
\nodata& $+$3.20 (0.37)& $+$2.57 (0.28)\\
$\mathrm{[O/Fe]} $&$<+$2.30 \nodata&$<+$2.90 \nodata&$<+$4.00 \nodata&$<+$2.50
\nodata&$<+$2.80 \nodata&$<+$1.70 \nodata\\
$\mathrm{[Na/Fe]}$&$>+$1.00 \nodata& $-$0.24 (0.15) &$<+$0.50 \nodata& $-$0.04
(0.13)& $+$0.19 (0.14)& $+$0.46(0.30)\\
$\mathrm{[Mg/Fe]}$& $+$0.64 (0.23) & $+$0.25 (0.14) & $+$0.59 (0.15) & $+$0.79
(0.14)& $+$0.42 (0.16)& $+$1.20(0.23)\\ 
$\mathrm{[Al/Fe]}$& $+$0.46 (0.24) & $-$0.38 (0.20) &$<+$0.03 \nodata& $-$0.57
(0.20)& $-$0.39 (0.21)& $-$0.38 (0.21)\\
$\mathrm{[Si/Fe]}$& \nodata \nodata& $+$0.05 (0.16) & $+$0.37 (0.15) & $+$0.80
(0.15)&$<+$0.25 \nodata& \nodata \nodata\\
$\mathrm{[Ca/Fe]}$& $+$0.49 (0.22) & $+$0.10 (0.13) & $+$0.34 (0.15) & $+$0.50
(0.14)& $+$0.00 (0.20)& $+$0.20 (0.22)\\
$\mathrm{[Sc/Fe]}$& \nodata \nodata& $+$0.00 (0.18) &$<+$0.70 \nodata& $-$0.12
(0.18)& $-$0.07 (0.16)& $+$0.09 (0.25)\\
$\mathrm{[Ti/Fe]}$& $+$0.71 (0.25) & $+$0.11 (0.21) & $+$0.18 (0.17) & $+$0.23
(0.14)& $+$0.35 (0.18)& $+$0.26 (0.22)\\
$\mathrm{[V/Fe]} $& $+$0.67 (0.22) &$<+$1.20 \nodata&$<+$2.50 \nodata&$<+$0.80
\nodata&$<+$0.90 \nodata& $+$0.30 (0.21)\\
$\mathrm{[Cr/Fe]}$& $+$0.08 (0.23) & $-$0.22 (0.18) &$<+$0.50 \nodata& $-$0.14
(0.17)& $-$0.49 (0.26)& $-$0.45 (0.22)\\
$\mathrm{[Mn/Fe]}$& $-$0.18 (0.24) & $-$0.65 (0.19) &$<+$0.20 \nodata& $-$0.60
(0.22)& $-$0.79 (0.20)& $-$0.96 (0.24)\\ 
$\mathrm{[Co/Fe]}$& $+$0.28 (0.25) & $+$0.20 (0.18) &$<+$1.60 \nodata& $+$0.30
(0.19)& $+$0.28 (0.16)& $+$0.30 (0.23)\\
$\mathrm{[Ni/Fe]}$& $+$0.37 (0.24) & $+$0.19 (0.19) &$<+$0.90 \nodata& $+$0.18
(0.17)& $-$0.12 (0.20)&$-$0.04 (0.29)\\
$\mathrm{[Zn/Fe]}$&$<+$0.90 \nodata&$<+$1.20 \nodata&$<+$2.60 \nodata&$<+$0.80
\nodata&$<+$1.30 \nodata&$<+$0.70 \nodata\\ 
$\mathrm{[Sr/Fe]}$& $+$0.25 (0.25) & $-$0.30 (0.19) & $+$0.32 (0.19) & $-$0.30
(0.19)& $-$1.08 (0.14)& $-$0.85 (0.20)\\
$\mathrm{[Ba/Fe]}$&$<-$1.80 \nodata&$<-$0.50 \nodata&$<+$0.80 \nodata& $-$1.27
(0.15)& $-$0.50 (0.15)& $-$0.90 (0.21)\\
$\mathrm{[Eu/Fe]}$&$<+$0.80 \nodata&$<+$1.50 \nodata&$<+$3.00 \nodata&$<+$1.50
\nodata&$<+$1.20 \nodata&$<+$0.50 \nodata\\
\enddata
\end{deluxetable*}

\begin{deluxetable*}{lrrrrr}
\tablecaption{Derived Abundances for CEMP-$s$ Stars\label{tab6}}
\tablewidth{0pt}
\tablehead{
\colhead{} & \colhead{\object{HE 0054$-$2542}} & \colhead{\object{HE 0440$-$3426}}
& \colhead{\object{HE 0450$-$4902}} & \colhead{\object{HE 1029$-$0546}} &
\colhead{\object{HE 2238$-$4131}}} 
\startdata
$\mathrm{[Fe/H]} $& $-$2.48 (0.20) & $-$2.19 (0.20) & $-$3.07 (0.20) & $-$3.28 (0.20)&$-$2.75 (0.20)\\
                  &        &        &        &       &        \\
A(Li)             &$<+$0.57 \nodata&$<+$0.26 \nodata&$<+$1.98 \nodata&$<+$2.00
\nodata&$<+$0.30 \nodata\\
                  &        &        &        &       &        \\
$\mathrm{[C/Fe]}$ & $+$2.13 (0.29) & $+$1.51 (0.25) & $+$2.03 (0.23) & $+$2.64 (0.20)&$+$2.63 (0.32)\\
$\mathrm{[N/Fe]}$ & $+$0.87 (0.27) & $+$0.78 (0.26) & $+$2.00 (0.29) & $+$2.90 (0.27)&$+$1.04 (0.33)\\
$\mathrm{[O/Fe]}$ &$<+$1.20 \nodata& $+$0.69 (0.17) &$<+$3.50 \nodata&$<+$3.70
\nodata&$<+$1.70 \nodata\\
$\mathrm{[Na/Fe]}$&$>+$1.20 \nodata& $+$0.67 (0.30) & $+$0.23 (0.22) & \nodata
\nodata&$>+$1.60 \nodata\\
$\mathrm{[Mg/Fe]}$& $+$0.78 (0.23) & $+$0.43 (0.21) & $+$0.53 (0.20) & $-$0.03 (0.17)&$+$0.87 (0.29)\\ 
$\mathrm{[Al/Fe]}$& $+$0.10 (0.21) & \nodata \nodata& $-$0.78 (0.26) &$<-$0.42
\nodata& $+$0.12 (0.28)\\
$\mathrm{[Si/Fe]}$& \nodata \nodata& \nodata \nodata& $+$0.00 (0.21) & $-$0.03 (0.18)&\nodata\nodata\\
$\mathrm{[Ca/Fe]}$& $+$0.40 (0.22) & $+$0.23 (0.20) & $+$0.70 (0.23) & $+$0.16 (0.20)&$+$0.43 (0.28)\\
$\mathrm{[Sc/Fe]}$& \nodata \nodata& \nodata \nodata& $+$0.12 (0.24) & $+$0.33 (0.21)&\nodata\nodata\\
$\mathrm{[Ti/Fe]}$& $+$0.42 (0.22) & $+$0.26 (0.20) & $+$0.58 (0.22) & $+$0.45 (0.20)&$+$0.44(0.28)\\
$\mathrm{[V/Fe]}$ & $+$0.28 (0.27) & $+$0.03 (0.19) &$<+$1.30 \nodata&$<+$1.20
\nodata& \nodata \nodata\\
$\mathrm{[Cr/Fe]}$& $-$0.03 (0.22) & $-$0.12 (0.20) & $+$0.03 (0.21) & $-$0.08 (0.18)&$+$0.00 (0.28)\\ 
$\mathrm{[Mn/Fe]}$& $-$0.38 (0.24) & $-$0.63 (0.22) & $-$0.73 (0.26) & $-$0.14 (0.24)&$-$0.63 (0.30)\\
$\mathrm{[Co/Fe]}$& $-$0.13 (0.25) & $-$0.54 (0.21) &$<+$0.60 \nodata& $+$0.90 (0.25)&$-$0.02 (0.29)\\
$\mathrm{[Ni/Fe]}$& $-$0.10 (0.27) & $+$0.03 (0.28) & $+$0.00 (0.20) & $+$0.34 (0.17)&$+$0.00 (0.34)\\
$\mathrm{[Zn/Fe]}$&$<+$0.08 \nodata& $+$0.06 (0.18) &$<+$1.40 \nodata&$<+$1.60
\nodata&$<+$0.70 \nodata\\ 
$\mathrm{[Sr/Fe]}$& $+$1.65 (0.20) & $+$0.33 (0.18) & $+$0.64 (0.26) & $+$0.07 (0.24)&$+$1.75 (0.27)\\
$\mathrm{[Y/Fe]}$ & $+$1.99 (0.23) & $+$0.33 (0.19) & \nodata \nodata&$<+$1.05
\nodata& $+$2.13 (0.28)\\
$\mathrm{[Zr/Fe]}$& $+$2.26 (0.22) & $+$0.64 (0.20) & \nodata \nodata&$<+$1.70
\nodata& $+$2.38 (0.29)\\
$\mathrm{[Ba/Fe]}$& $+$1.52 (0.26) & $+$0.46 (0.19) & $+$1.21 (0.20) & $+$0.80 (0.17)&$+$1.80 (0.28)\\
$\mathrm{[La/Fe]}$& $+$1.63 (0.24) & $+$1.18 (0.20) & \nodata \nodata& \nodata
\nodata& $+$2.32 (0.28)\\
$\mathrm{[Ce/Fe]}$& $+$1.50 (0.23) & $+$0.89 (0.18) & \nodata \nodata&$<+$2.70
\nodata& $+$2.35 (0.27)\\
$\mathrm{[Pr/Fe]}$& $+$1.60 (0.23) & $+$1.07 (0.20) & \nodata \nodata& \nodata
\nodata& $+$2.26 (0.27)\\
$\mathrm{[Nd/Fe]}$& $+$0.36 (0.23) & $+$0.30 (0.17) & \nodata \nodata&$<+$2.46
\nodata& $+$1.05 (0.26)\\
$\mathrm{[Sm/Fe]}$& $+$1.33 (0.21) & $+$1.01 (0.24) & \nodata \nodata& \nodata
\nodata& $+$1.70 (0.31)\\
$\mathrm{[Eu/Fe]}$& $+$0.78 (0.23) &$<+$0.62 \nodata&$<+$2.00 \nodata&$<+$2.50
\nodata& $+$1.10 (0.29)\\
$\mathrm{[Gd/Fe]}$& $+$1.10 (0.24) & \nodata \nodata& \nodata \nodata& \nodata
\nodata& \nodata \nodata\\
$\mathrm{[Dy/Fe]}$& $+$1.20 (0.24) & $+$0.74 (0.21) & \nodata \nodata& \nodata
\nodata& $+$1.70 (0.29)\\
$\mathrm{[Er/Fe]}$& \nodata \nodata& $+$1.14 (0.21) & \nodata \nodata& \nodata
\nodata&$>+$2.00 \nodata\\
$\mathrm{[Pb/Fe]}$&$<+$1.50 \nodata& $+$1.64 (0.23) &$<+$3.00 \nodata& $+$3.34
(0.23)&$<+$2.00 \nodata 
\enddata
\end{deluxetable*}

\begin{deluxetable*}{lrrrr}
\tablecaption{Derived Abundances for CEMP-$r$ and CEMP-$r/s$ Stars\label{tab7}}
\tablewidth{0pt}
\tablehead{\colhead{}&\multicolumn{2}{c}{CEMP-$r$}&\multicolumn{2}{c}{CEMP-$r/s$}\\
\hline
\colhead{}&\multicolumn{2}{c}{}&\multicolumn{2}{c}{}\\
\colhead{} & \colhead{\object{HE 0010$-$3422}} & \colhead{\object{HE 0448$-$4806}}
& \colhead{\object{HE 0243$-$3044}} & \colhead{\object{HE 2208$-$1239}} }
\startdata
$\mathrm{[Fe/H]} $& $-$2.78 (0.20)& $-$2.26 (0.20)& $-$2.58 (0.20)& $-$2.88 (0.20)\\
                  &        &        &        &       \\
A(Li)             &$<+$1.11 \nodata&$<+$1.49 \nodata&$<+$0.97 \nodata&$<+$0.77
\nodata\\
                  &        &        &        &       \\
$\mathrm{[C/Fe]}$ & $+$1.92 (0.31) & $+$2.24 (0.29) & $+$2.43 (0.27) & $+$1.30 (0.29)\\
$\mathrm{[N/Fe]}$ & $+$2.60 (0.27) & $+$1.44 (0.29) & $+$1.48 (0.28) & $+$1.95 (0.25)\\
$\mathrm{[O/Fe]}$ &$<+$2.02 \nodata&$<+$1.90 \nodata&$<+$1.90 \nodata&$<+$1.40 \nodata\\
$\mathrm{[Na/Fe]}$& $+$1.00 (0.28) &$>+$0.70 \nodata&$>+$1.00 \nodata& \nodata \nodata\\
$\mathrm{[Mg/Fe]}$& $+$0.34 (0.23) & $+$0.44 (0.20) & $+$1.08 (0.23) & $+$0.59 (0.20)\\ 
$\mathrm{[Al/Fe]}$& $-$0.54 (0.21) & $-$0.23 (0.24) & $+$0.04 (0.22) & $-$0.32 (0.18)\\
$\mathrm{[Si/Fe]}$& \nodata \nodata& \nodata \nodata& \nodata \nodata& \nodata \nodata\\
$\mathrm{[Ca/Fe]}$& $+$0.26 (0.22) & $+$0.35 (0.20) & $+$0.12 (0.22) & $+$0.45 (0.19)\\
$\mathrm{[Sc/Fe]}$& $+$0.45 (0.24) & $+$0.25 (0.23) & \nodata \nodata& $+$0.25 (0.19)\\
$\mathrm{[Ti/Fe]}$& $+$0.59 (0.22) & $+$0.50 (0.20) & $+$0.43 (0.23) & $+$0.70 (0.19)\\
$\mathrm{[V/Fe]} $& $+$0.73 (0.27) & $+$0.38 (0.21) & $+$0.73 (0.22) & $+$0.54 (0.25)\\
$\mathrm{[Cr/Fe]}$& $-$0.23 (0.25) & $-$0.03 (0.22) & $-$0.07 (0.23) & $-$0.20 (0.22)\\ 
$\mathrm{[Mn/Fe]}$& $-$0.71 (0.24) & $-$0.37 (0.26) & $-$0.47 (0.25) & $-$0.83 (0.22)\\
$\mathrm{[Co/Fe]}$& $+$0.15 (0.25) & $+$0.04 (0.25) & $+$0.03 (0.24) & $+$0.10 (0.22)\\
$\mathrm{[Ni/Fe]}$& $+$0.01 (0.27) & $+$0.14 (0.22) & $+$0.26 (0.30) & $-$0.10 (0.24)\\
$\mathrm{[Zn/Fe]}$& $+$0.57 (0.21) & $+$0.33 (0.24) &$<+$0.50 \nodata& $+$0.36 (0.18)\\ 
$\mathrm{[Sr/Fe]}$& $+$0.85 (0.20) & $+$1.10 (0.23) & $+$0.97 (0.21) & $+$0.50 (0.17)\\
$\mathrm{[Y/Fe]} $& $+$1.01 (0.23) & $+$0.93 (0.22) & $+$0.99 (0.22) & $+$0.37 (0.20)\\
$\mathrm{[Zr/Fe]}$& $+$1.08 (0.22) & $+$1.10 (0.21) & $+$1.06 (0.23) & $+$0.84 (0.19)\\
$\mathrm{[Ba/Fe]}$& $+$1.54 (0.26) & $+$1.78 (0.20) & $+$1.96 (0.22) & $+$1.68 (0.23)\\
$\mathrm{[La/Fe]}$& $+$2.21 (0.24) & $+$2.33 (0.23) & $+$2.51 (0.23) & $+$1.96 (0.21)\\
$\mathrm{[Ce/Fe]}$& $+$1.99 (0.23) & $+$2.20 (0.23) & $+$2.32 (0.21) & $+$1.80 (0.21)\\
$\mathrm{[Pr/Fe]}$& $+$2.00 (0.20) & $+$2.24 (0.22) & $+$2.48 (0.22) & $+$1.77 (0.20)\\
$\mathrm{[Nd/Fe]}$& $+$1.30 (0.23) & $+$1.46 (0.22) & $+$1.69 (0.20) & $+$1.06 (0.20)\\
$\mathrm{[Sm/Fe]}$& $+$1.97 (0.21) & $+$2.09 (0.21) & $+$2.18 (0.26) & $+$1.76 (0.18)\\
$\mathrm{[Eu/Fe]}$& $+$1.72 (0.23) & $+$1.87 (0.23) & $+$1.90 (0.24) & $+$1.52 (0.21)\\
$\mathrm{[Gd/Fe]}$& \nodata \nodata& $+$1.92 (0.24) & $+$2.35 (0.23) & $+$1.61 (0.22)\\
$\mathrm{[Dy/Fe]}$& \nodata \nodata& $+$1.89 (0.23) & $+$1.80 (0.24) &\nodata \nodata\\
$\mathrm{[Er/Fe]}$& \nodata \nodata& $+$2.78 (0.22) & $+$2.64 (0.24) & $+$1.97 (0.20)\\
$\mathrm{[Pb/Fe]}$ &$+$2.62 (0.27) & $+$3.17 (0.29) & $+$3.07 (0.25) & $+$1.70 (0.36)
\enddata
\end{deluxetable*}

\begin{deluxetable}{lrc}
\tablecaption{$^{12}$C/$^{13}$C Isotopic Ratios\label{tab8}}
\tablewidth{0pt}
\tablehead{\colhead{Star} & \colhead{$^{12}$C/$^{13}$C} & \colhead{Type}}
\startdata
\object{HE 0010-3422} &   5 & CEMP-$r$\\
\object{HE 0054-2542} &  16 & CEMP-$s$\\ 
\object{HE 0100-1622} &  13 & CEMP-$no$\\
\object{HE 0134-1519} &$>$4 & CEMP-$no$\\
\object{HE 0233-0343} &$>$5 & CEMP-$no$\\
\object{HE 0243-3044} &  10 & CEMP-$r/s$\\
\object{HE 0440-3426} &  13 & CEMP-$s$\\ 
\object{HE 0448-4806} &  10 & CEMP-$r$\\
\object{HE 1029-0546} &   9 & CEMP-$s$\\
\object{HE 1310-0536} &   3 & CEMP-$no$\\ 
\object{HE 2208-1239} &   4 & CEMP-$r/s$\\  
\object{HE 2238-4131} &  16 & CEMP-$s$\\ 
\object{HE 2331-7155} &   5 & CEMP-$no$\\ 
\enddata
\end{deluxetable}

As many of our sample stars are carbon enhanced, resulting in
blended lines, we have chosen to perform spectral synthesis to derive
the abundances of individual elements, whereas abundances for the
comparison sample of \citet{yong2013} were derived mostly from
equivalent-width measurements. All other aspects of the analysis are
the same for the two samples, thus, by combining them we have a sample
of over 200 homogeneously analyzed metal-poor stars. As a
demonstration, we have derived abundances for HE~0146$-$1548 from the
sample of \cite{yong2013}. These are listed in Table~\ref{tabcom} along
with the abundances listed in \cite{yong2013}. As can be seen, the two
sets of abundances derived for this star agree very well.

The combined sample includes examples of all the different stellar
abundance patterns that are commonly found at very low metallicity, for
both carbon-enhanced stars (all four sub-classes are represented) and
non carbon-enhanced stars. With this variety, and the fact that the
combined sample includes some of the most metal-poor stars known, we are
able to carry out a more detailed investigation of the signatures of our
Galaxy's early chemical evolution than previous possible.

\begin{deluxetable}{lrr}
\tablecaption{Abundances Derived for HE~0145$-$1548 \label{tabcom}}
\tablewidth{0pt}
\tablehead{\colhead{$\mathrm{[X/Fe]}$} & \colhead{\cite{yong2013}} &
  \colhead{This work}}
\startdata
$\mathrm{[C/Fe]}$  & $+$0.84 & $+$0.80\\
$\mathrm{[Na/Fe]}$ & $+$1.17 & $+$1.15\\
$\mathrm{[Mg/Fe]}$ & $+$0.87 & $+$0.86\\
$\mathrm{[Al/Fe]}$ & $+$0.14 & $+$0.10\\
$\mathrm{[Si/Fe]}$ & $+$0.50 &$<+$0.66\\
$\mathrm{[Ca/Fe]}$ & $+$0.22 & $+$0.19\\
$\mathrm{[Ti/Fe]}$ & $+$0.17 & $+$0.15\\
$\mathrm{[Cr/Fe]}$ & $-$0.38 & $-$0.39\\
$\mathrm{[Mn/Fe]}$ & $-$0.59 & $-$0.64\\
$\mathrm{[Co/Fe]}$ & $+$0.30 & \nodata\\
$\mathrm{[Ni/Fe]}$ & $+$0.05 & $+$0.07\\
$\mathrm{[Sr/Fe]}$ & $-$0.38 & $-$0.34\\
$\mathrm{[Ba/Fe]}$ & $-$0.71 & $-$0.78\\
\enddata
\end{deluxetable}

\cite{yong2013} did not present Ba and Eu abundances for all the stars in their
sample. However, abundances of these two elements are essential for the
classification of CEMP stars (CEMP-$no$: $\mathrm{[Ba/Fe]} < 0$,
CEMP-$s$: $\mathrm{[Ba/Fe]} > +1$, $\mathrm{[Ba/Eu]} > +0.5$, CEMP-$r$:
$\mathrm{[Eu/Fe]} > +1$, CEMP-$r/s$: $0.0 < \mathrm{[Ba/Eu]} < +0.5$ ),
thus we have searched the literature for Ba and Eu abundances for the
full sample of CEMP stars from \cite{yong2013}; these additional
abundances are listed in Table~\ref{tab9}. The supplemental abundances
are only used for classification of the CEMP stars, and not included in
the plots, as they have not been derived in the same homogeneous manner
as the abundances presented here. Table~\ref{tab9} also includes upper
limits for Ba in HE~0107$-$5240, HE~0557$-$4840, and HE~1327$-$2326, as
these all lie in the sparsely populated region with
$\mathrm{[Fe/H]}<-4.0$; we include them in our plots. The combined
sample includes 143 NMP, 32 CEMP-$no$, 30 CEMP-$s$, 4 CEMP-$r$, and 4
CEMP-$r/s$ stars.

\begin{deluxetable}{lrc}
\tablecaption{Ba and Eu Abundances from the Literature\label{tab9}}
\tablewidth{0pt}
\tablehead{\colhead{Star} & \colhead{$\mathrm{[X/Fe]}$} & \colhead{Ref}}
\startdata
\cutinhead{Ba}
\object{BD-18:5550}   &  $-$0.74 & \citet{francois2007}\\
\object{CS 22880$-$074} &  $+$1.31 & \citet{aoki2002b}\\
\object{CS 22892$-$052} &  $+$0.99 & \citet{sneden2003}\\
\object{CS 22897$-$008} &  $-$1.00 & \citet{francois2007}\\
\object{CS 29498$-$043} &  $-$0.45 & \citet{aoki2002a}\\
\object{CS 29516$-$024} &  $-$0.90 & \citet{francois2007}\\
\object{CS 30301$-$015} &  $+$1.45 & \citet{aoki2002b}\\
\object{CS 31062$-$050} &  $+$2.30 & \citet{aoki2002b}\\
\object{HD 196944}    &  $+$1.10 & \citet{aoki2002b}\\
\object{HE 0107$-$5240} & $<+$0.82 & \citet{christlieb2004}\\
\object{HE 0557$-$4840} & $<+$0.03 & \citet{norris2007}\\
\object{HE 1300$+$0157} & $<-$0.63 & \citet{cohen2008}\\
\object{HE 1327$-$2326} & $<+$1.46 & \citet{aoki2006}\\
\cutinhead{Eu}
\object{CS 22880$-$074} &  $+$0.50 & \citet{aoki2002b}\\
\object{CS 22892$-$052} &  $+$1.64 & \citet{sneden2003}\\
\object{CS 22948$-$027} &  $+$1.57 & \citet{aoki2002b}\\
\object{CS 29497$-$034} &  $+$1.80 & \citet{barbuy2005}\\
\object{CS 29503$-$010} &  $+$1.69 & \citet{allen2012}\\
\object{CS 30301$-$015} &  $+$0.20 & \citet{aoki2002b}\\
\object{CS 31062$-$012} &  $+$1.62 & \citet{aoki2002b}\\
\object{CS 31062$-$050} &  $+$1.84 & \citet{aoki2002b}\\
\object{HD 196944}    &  $+$0.17 & \citet{aoki2002b}\\
\object{HE 0143$-$0441} &  $+$1.46 & \citet{cohen2006}\\
\object{HE 0336$+$0113} &  $+$1.18 & \citet{cohen2013}\\
\object{HE 1031$-$0020} & $<+$0.87 & \citet{cohen2006}\\
\object{HE 2158$-$0348} &  $+$0.80 & \citet{cohen2006}\\
\enddata
\end{deluxetable}

\subsection{Carbon and Nitrogen}

Carbon and nitrogen are among the first elements to be synthesized
in the universe following the Big Bang. Yet our understanding of the
abundances of these two elements found in the most metal-poor stars of
our Galaxy is still limited. It is recognized that the observed carbon
and nitrogen abundances of a given star are subject to change as the
star evolves past the main sequence and up the red giant branch. Material that
has been C-depleted and N-enhanced in the lower layers of a stellar atmosphere
is transported to the surface of the star via mixing, enhancing N at the
expense of C at the surface of the star. Thus, if we require the abundances of
these two elements at the time when the star was born, in order to better
constrain its progenitor, we need to correct the abundances of C and N for
these evolutionary effects.

\cite{placco2014b} have developed a method for correcting the C
abundances in metal-poor stars according to the evolutionary state of
the star, based on the STARS stellar evolution code \citep{eggleton1971,
stancliffeeldridge2009} and thermohaline mixing as described in
\cite{stancliffe2009}. We have corrected the C abundances for the giant
stars in our sample using this method; for the \cite{yong2013} sample we
have used the corrections listed in \cite{placco2014b}, while corrections
for our own sample can be seen in Table~\ref{tab10} (only including the
non-zero corrections). For the remainder of this work we employ the
corrected C abundances, unless stated otherwise. Note that no explicit correction is currently
made for N.

\begin{deluxetable*}{lccccc}
\tablecaption{Carbon Abundances Corrected for Stellar-Evolution Effects\label{tab10}}
\tablewidth{0pt}
\tablehead{
\colhead{Star} & \colhead{$\log$g} &
\colhead{$\mathrm{[Fe/H]}$} & \colhead{$\mathrm{[C/Fe]}_{original}$} &
\colhead{$\mathrm{[C/Fe]}_{corrected}$} & \colhead{$\mathrm{[N/Fe]}$}}
\startdata
HE~0054$-$2542 &2.69 & $-$2.48 & $+$2.13 & $+$2.15 & $+$0.87 \\
HE~0440$-$3426 &1.56 & $-$2.19 & $+$1.51 & $+$1.64 & $+$0.78 \\
HE~1310$-$0536 &1.85 & $-$4.15 & $+$2.36 & $+$2.47 & $+$3.20 \\   	
HE~1429$-$0347 &1.92 & $-$2.71 & $+$0.31 & $+$0.45 & $+$1.89 \\
HE~2159$-$0551 &1.46 & $-$2.81 & $-$0.24 & $+$0.22 & $+$0.88 \\
HE~2208$-$1239 &2.32 & $-$2.88 & $+$1.30 & $+$1.31 & $+$1.95 \\
HE~2238$-$4131 &2.53 & $-$2.75 & $+$2.63 & $+$2.65 & $+$1.04 \\
HE~2331$-$7155 &1.54 & $-$3.68 & $+$1.34 & $+$1.69 & $+$2.57
\enddata
\end{deluxetable*}

\subsubsection{$\mathrm{[C/N]}>0$ vs. $\mathrm{[C/N]}<0$ stars}

Many of the carbon-enhanced stars are also found to be enhanced in
nitrogen. However, the minority of VMP, EMP, and UMP stars have both
elements detected, a deficiency that surely needs to be addressed in the
near future. Only 79 of the 193 stars in the sample of \cite{yong2013}
have detections of both C and N. In our sample of 23 stars, we detect C
and N simultaneously for as many as 14 stars, increasing the number of
stars with C and N detections in the combined sample by about 20\%. In
the combined sample we have 104 stars with both C and N detections, plus
a few with a detection of either C or N and an upper limit on the other.
Roughly equal numbers of these are NMP and CEMP stars. 

\begin{figure}
\includegraphics[angle=0,width=3.5in]{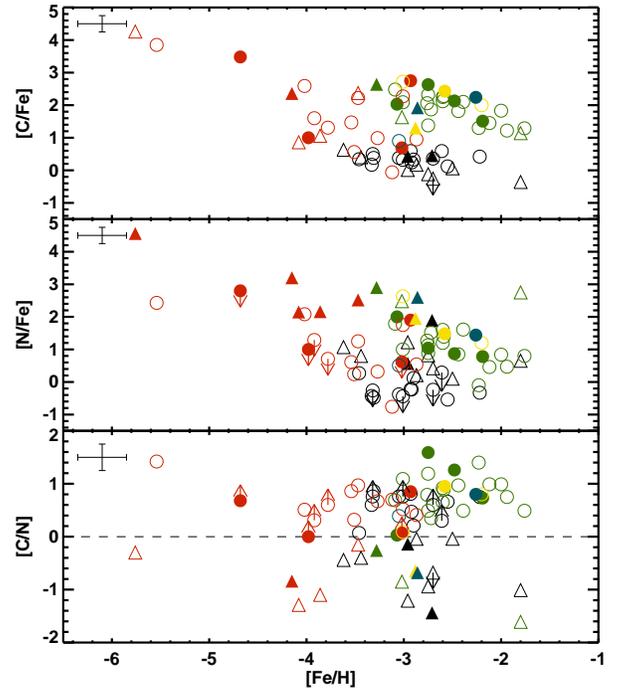}
\caption{C and N abundances and $\mathrm{[C/N]}$ ratios for stars in this sample (filled symbols) and that of
  \citet{yong2013} (unfilled symbols). Circles represent stars with
$\mathrm{[C/N]}>0$; triangles are stars with
$\mathrm{[C/N]}<0$. Symbols are color-coded according to
black: ``normal'' metal-poor (NMP) stars, red: CEMP-$no$ stars, green:
CEMP-$s$ stars, blue: CEMP-$r$ stars, yellow: CEMP-$r/s$ stars. An
approximate error bar for the sample stars is shown in the upper left of
each panel. \label{fig3}} 
\end{figure}

Figure~\ref{fig3} shows the C and N abundances and the
$\mathrm{[C/N]}$ ratios of our combined sample of stars as a function
of metallicity. We have divided the stars into two groups -- dots
indicate stars where $\mathrm{[C/N]}>0$
($\mathrm{[C/Fe]}>\mathrm{[N/Fe]}$) and triangles indicate stars where
$\mathrm{[C/N]}<0$ ($\mathrm{[C/Fe]}<\mathrm{[N/Fe]}$). As we do not
have the N abundance corrections corresponding to the C corrections
mentioned above, this plot only includes dwarfs, subgiants, and early
giants, for which the corrections in C and N will not alter the
$\mathrm{[C/N]}>0$ or $\mathrm{[C/N]}<0$ status of the star. For the
remainder of this paper we refer to this subsample as the ``C/N stars''.  

\begin{figure}
\includegraphics[angle=0,width=3.5in]{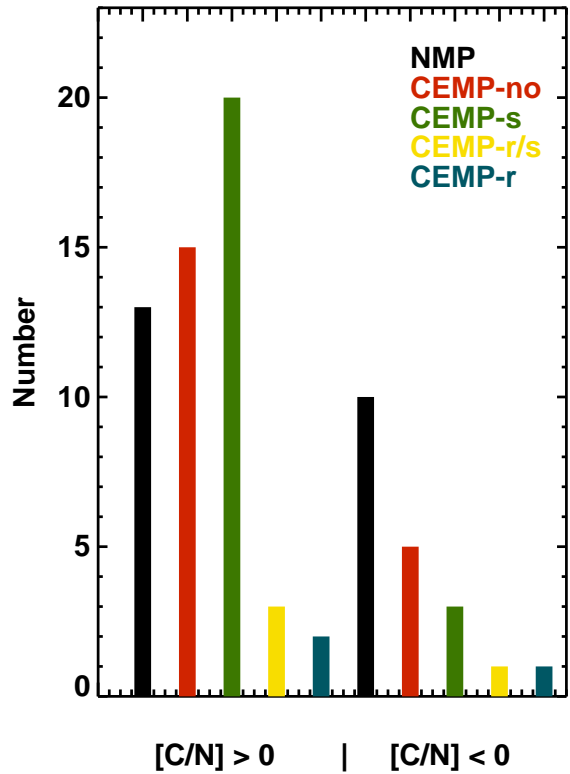}
\caption{Number of stars with either $\mathrm{[C/N]}>0$
or $\mathrm{[C/N]}<0$ for the different types of metal-poor
stars. Black: NMP stars, red: CEMP-$no$ stars, green: CEMP-$s$ stars, blue: CEMP-$r$ stars, 
yellow: CEMP-$r/s$ stars\label{fig4}} 
\end{figure}

Figure~\ref{fig4} shows the number of each type of metal-poor star with
either $\mathrm{[C/N]}>0$ or $\mathrm{[C/N]}<0$. There are clearly more
stars with $\mathrm{[C/N]}>0$ than with $\mathrm{[C/N]}<0$. For the NMP
stars the numbers are roughly equal, while the CEMP-$no$ and especially
the CEMP-$s$ stars are of the $\mathrm{[C/N]}>0$ variety. From
inspection of the bottom panel of Figure~\ref{fig3}, none of the
CEMP-$no$ stars with $\mathrm{[C/N]}<0$ are found with metallicities
above $\mathrm{[Fe/H]}>-3.4$; all CEMP-$no$ stars with
$\mathrm{[C/N]}<0$ are at the extremely low-metallicity end.

\begin{figure}
\includegraphics[angle=0,width=3.5in]{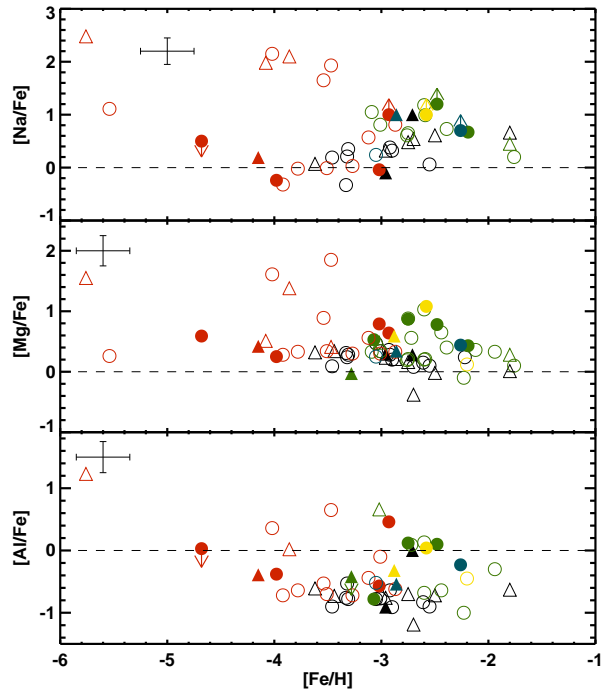}
\caption{$\mathrm{[Na/Fe]}$, $\mathrm{[Mg/Fe]}$, and $\mathrm{[Al/Fe]}$ for the
C/N stars. Color-coding of the stellar classes is as in Figure~\ref{fig3}. An
approximate error bar for the sample stars is shown in the upper left of
each panel.  \label{fig5}} 
\end{figure}

\begin{figure}
\includegraphics[angle=0,width=3.5in]{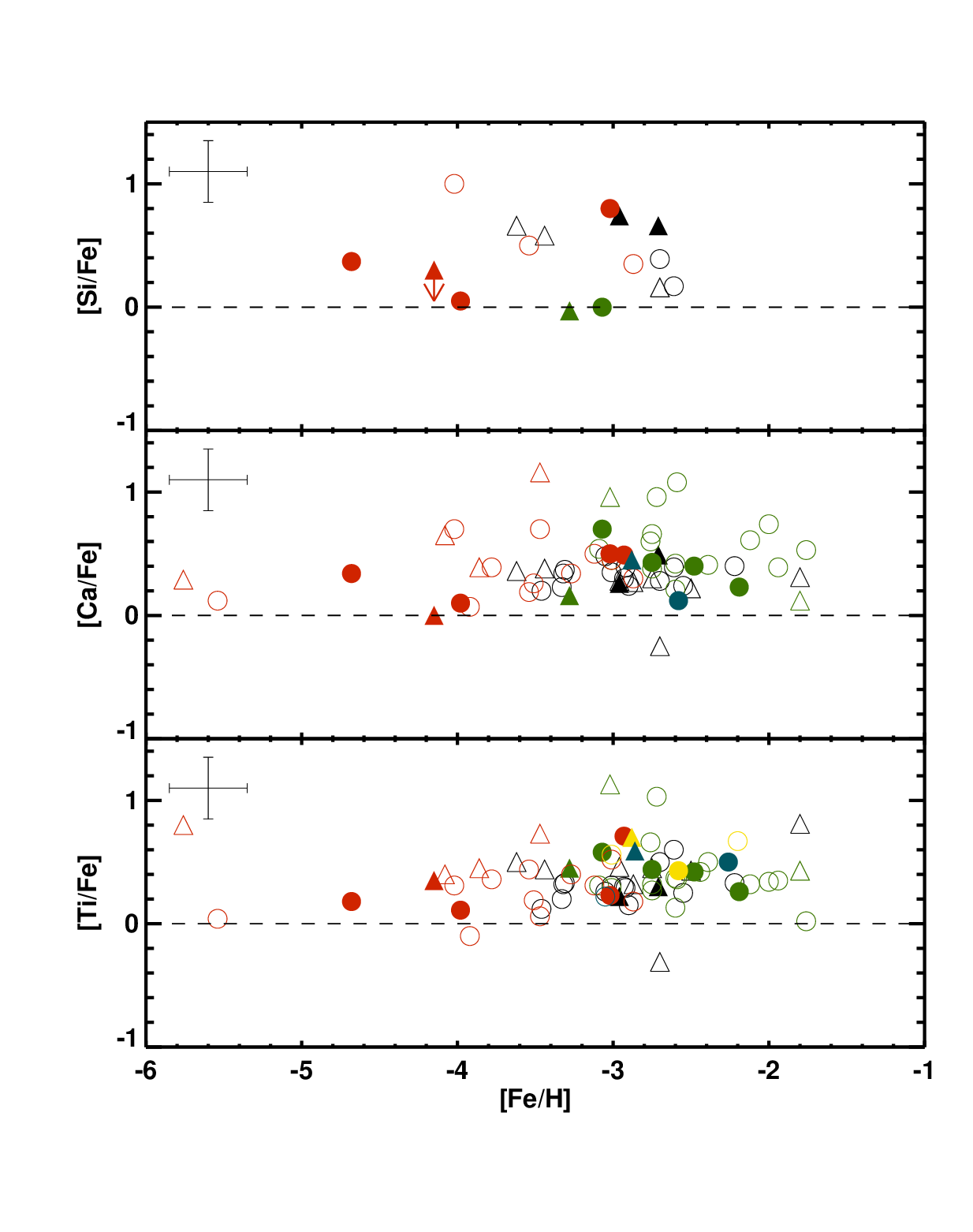}
\caption{$\mathrm{[Si/Fe]}$, $\mathrm{[Ca/Fe]}$, and $\mathrm{[Ti/Fe]}$  for the
C/N stars. Color-coding of the stellar classes is as in Figure~\ref{fig3}. An
approximate error bar for the sample stars is shown in the upper left of
each panel.  \label{fig6}} 
\end{figure}

\begin{figure}
\includegraphics[angle=0,width=3.5in]{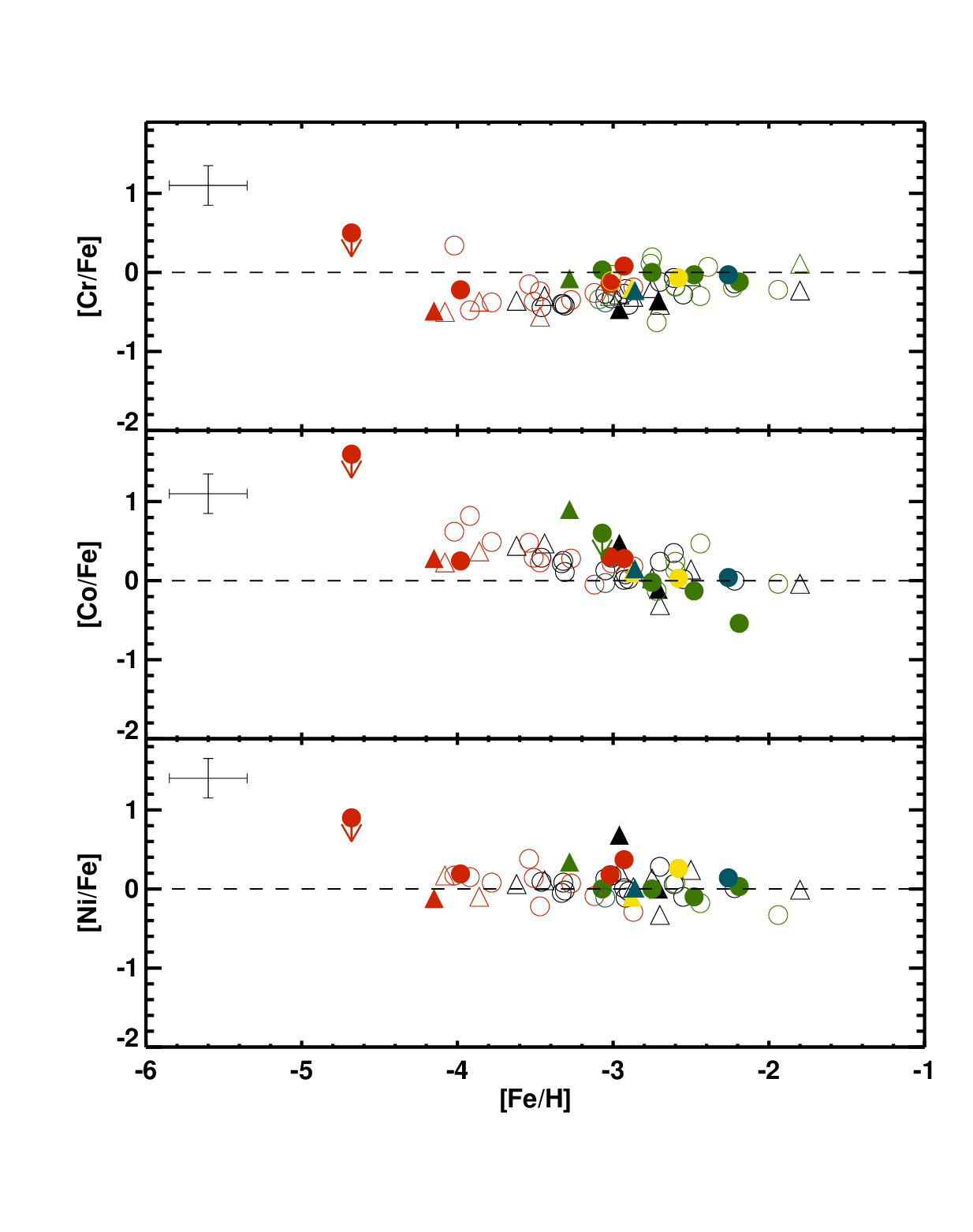}
\caption{$\mathrm{[Cr/Fe]}$, $\mathrm{[Co/Fe]}$, and
$\mathrm{[Ni/Fe]}$ for the C/N stars. Color-coding of the stellar classes 
is as in Figure~\ref{fig3}. An approximate error bar for the sample
stars is shown in the upper left of each panel. \label{fig7}} 
\end{figure}

\begin{figure}
\includegraphics[angle=0,width=3.5in]{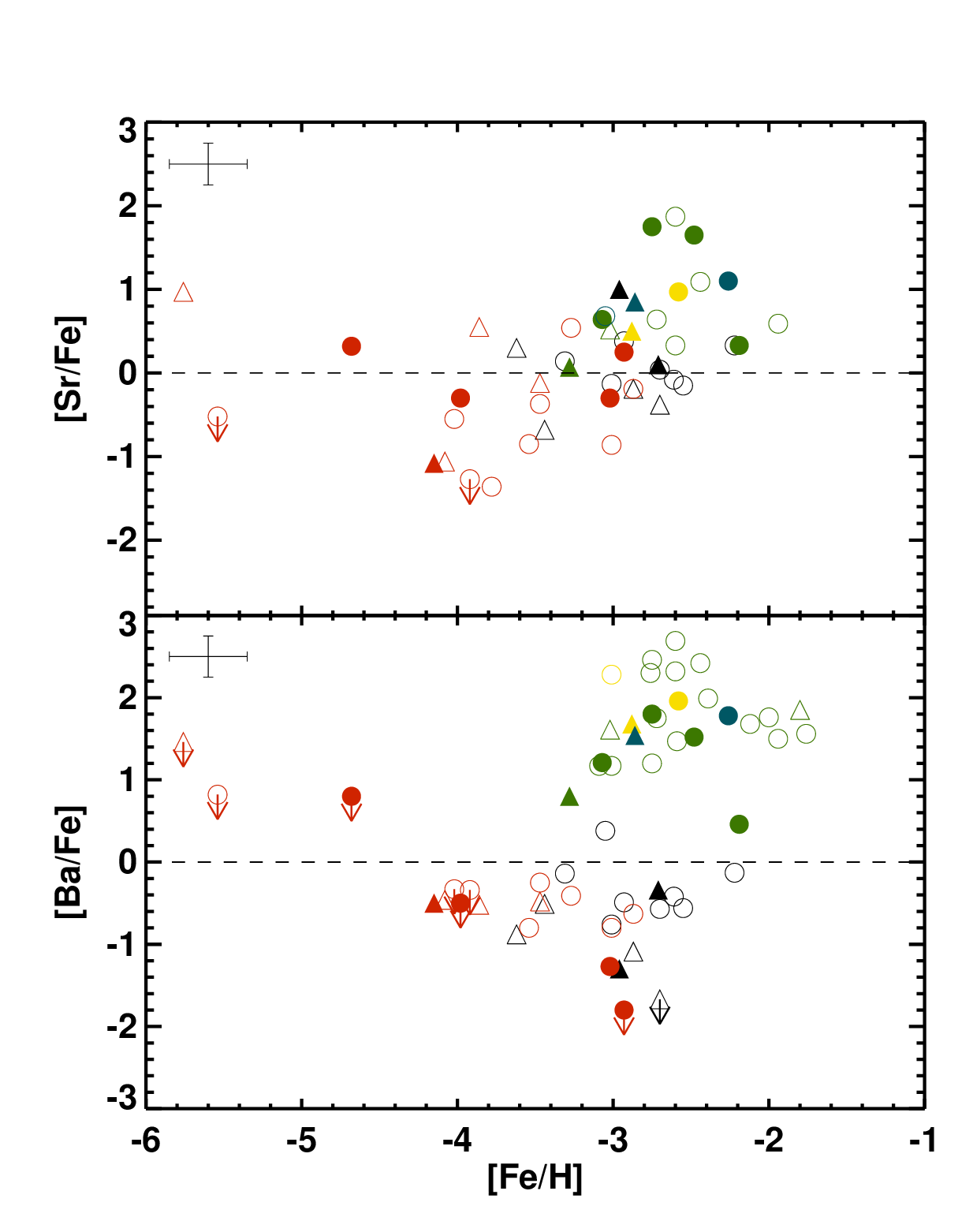}
\caption{$\mathrm{[Sr/Fe]}$ and $\mathrm{[Ba/Fe]}$ for the C/N stars. 
Color-coding of the stellar classes is as in Figure~\ref{fig3}. An
approximate error bar for the sample stars is shown in the upper left of
each panel. \label{fig8}} 
\end{figure}

Examination of the abundance ratios for the remaining elements for these
stars using this division between C- and N-dominated stars, plotted in
Figure~\ref{fig5}, Figure~\ref{fig6}, Figure~\ref{fig7}, and
Figure~\ref{fig8}, indicates that $\mathrm{[Fe/H]}=-3.4$ also serves as
a dividing line for the Na and Mg abundances in CEMP-$no$ stars. Below
this metallicity, stars with large over-abundances of these two elements
appear, in contrast to the behavior above this metallicity. Two of the
CEMP-$no$ stars with $\mathrm{[C/N]}<0$, HE~1327$-$2326 and
HE~2323$-$0256, show large enhancements in Na, Mg and Sr, while the
other two stars, HE~1150$-$0428 and HE~1310$-$0536, both have
$\mathrm{[Mg/Fe]}\sim0.4$ and subsolar $\mathrm{[Sr/Fe]}$.

For the remainder of the elements we see the well-known abundance
patterns found at low metallicity in these plots -- moderate
over-abundances of the $\alpha$-elements and very low star-to-star
scatter for both the $\alpha$- and iron-peak elements, with a large
scatter for the neutron-capture elements.

An additional sub-class of metal-poor stars has been defined for
stars with $\mathrm{[N/Fe]} > +0.5$ and $\mathrm{[C/N]} < -0.5$, the
so-called nitrogen-enhanced metal-poor (NEMP) stars \citep{johnson2007}.
These stars are expected to be the result of mass transfer from an
intermediate-mass AGB star that has undergone hot bottom burning, and
thereby produced large amounts of nitrogen. Very few of these stars are
known to exist, so here we consider the six stars in our sample
with $\mathrm{[C/N]} < 0$ to see if they are possible NEMP stars.
Two of the stars, HE~1310$-$0536 and HE~2331$-$7155, are CEMP-$no$
stars, and two, HE~1429$-$0347 and HE~2159$-$0551, are NMP stars. None of
these exhibits excesses of neutron-capture elements, as would be expected
for the NEMP stars, thus they can be excluded from this class. This
leaves HE~0010$-$3422 (a CEMP-$r$ star) and HE~2208$-$1239 (a CEMP-$r/s$
star) as NEMP candidates. The binary status of these two stars is not
known, but future radial-velocity monitoring should be able to clarify
if they are consistent with an NEMP classification.

\subsubsection{The Carbon Plateau(s)} 

\cite{spite2013} suggested the presence of two separate plateaus or 
``bands'' in the distribution of C abundances, as a function of
$\mathrm{[Fe/H]}$, for VMP and EMP stars. The C abundances for stars
with metallicities $\mathrm{[Fe/H]}> -3$ appeared to cluster around the
solar carbon abundance ($A(\rm C) \sim 8.5$\footnote{Here we employ the
standard notation that $A(\rm X) = \log\epsilon\,(X) + 12.0$.}), while
those with $\mathrm{[Fe/H]} < -3$ (including the lowest metallicity stars
known) cluster around a lower C abundance, $A(\rm C) \sim 6.5$. These
authors proposed that the two bands could be associated with differing
astrophysical production sites for the C in these stars -- those in the
higher band being the result of mass transfer of C from an asymptotic
giant-branch (AGB) companion (i.e., extrinsic enrichment), whereas those
in the lower band being the result of C that is intrinsic to the star
(that is, the C was already present in the ISM from which the star was
born). It is useful to note that \cite{spite2013} only used dwarfs and
turnoff stars in their study, stars where the C abundances are not
expected to be altered due to evolutionary effects. The recent paper
by \cite{bonifacio2015} confirms the existence of the two carbon
bands for a larger sample, including the stars from \cite{yong2013}.

\begin{figure}
\includegraphics[angle=0,width=3.5in]{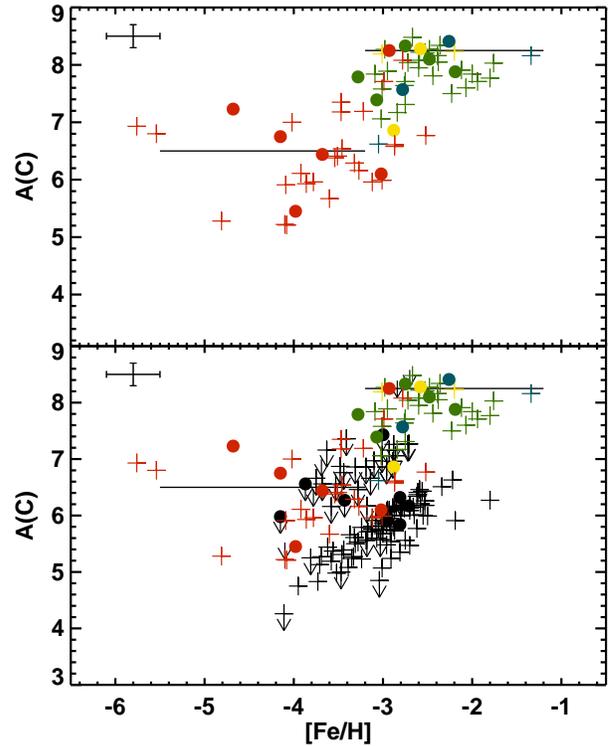}
\caption{$A(\rm C)$ abundances for sample stars (dots) and stars from \cite{yong2013}
(plusses). The top panel shows only CEMP stars, while all stars
(including non carbon-enhanced stars) are shown in the bottom panel. The
two carbon bands are indicated by solid lines, and upper limits on
individual C abundances are indicated by arrows.  Symbols are color-coded according to
black: ``normal'' metal-poor (NMP) stars, red: CEMP-$no$ stars, green:
CEMP-$s$ stars, blue: CEMP-$r$ stars, yellow: CEMP-$r/s$ stars. An
approximate error bar for the sample stars is shown in the upper left of
each panel. 
\label{fig9}} 
\end{figure}

We consider this question again with our new sample. Figure~\ref{fig9}
shows the absolute carbon abundances, $A(\rm C)$, for the stars in our
new sample (dots) along with those of \citet{yong2013} (plusses), as a
function of $\mathrm{[Fe/H]}$. The top panel shows only the CEMP stars,
while the bottom panel shows both the CEMP and NMP stars. There does
indeed appear to exist a difference in the C abundances for the lower
metallicity and higher metallicity CEMP stars, as suggested by
\cite{spite2013}. Our larger dataset exhibits a smoother transition
between the two bands in the metallicity region $\mathrm{[Fe/H]}\sim
-3.4$ to $\mathrm{[Fe/H]} \sim -3.2$. \cite{bonifacio2015} identify
four CEMP-$no$ stars with C abundances on the high carbon band. In
Figure~\ref{fig9} it can be seen that our sample includes three
CEMP-$no$ stars with C abundances on the high band. These three stars --
HE~0100$-$1622 ($\mathrm{[Ba/Fe]} < -1.80$), HE~2202$-$4831
($\mathrm{[Ba/Fe]} = -1.28$) and HE~2356$-$0410 ($\mathrm{[Ba/Fe]} =
-0.80$) are all confirmed CEMP-$no$ stars, hence they challenge the
interpretation of the two bands as being solely due to extrinsic and
intrinsic processes. We note that the binary status of this handful of
stars is not presently known, and would clearly be of great interest to
constrain. If the CEMP-$no$ stars found on the high carbon band are
indeed the result of mass transfer in a binary system, it will be
difficult to explain how large amounts of carbon but no or very small
amounts of $s$-process elements have been transferred from their AGB
companion. 

We also see that CEMP-$r$ and CEMP-$r/s$ stars are found at both the
high and the low levels of C enhancement. It would be interesting to
examine larger samples of these stars, in order to search for the
possible dominance of either high or low carbon-abundance stars for
either of these CEMP sub-classes.

The large carbon enhancements found in the lowest metallicity stars
is expected to be related to the formation of low-mass stars in the
early universe. It has been demonstrated that low-mass stars can form as
a result of cooling of gas clouds via fine-structure lines of carbon and
oxygen \citep{bromm2003,frebel2007}. Hence, the large C abundances found
at the lowest metallicities in our sample support the formation of
low-mass stars via this channel.

\subsection{$^{12}$C/$^{13}$C Isotopic Ratios}

When internal mixing occurs in stars, whether that mixing is due to
convection driven by rapid rotation (``spinstars,'' see
\citealt{meynet2006,hirschi2007,maeder2014}) or convection in AGB stars
during their evolution \citep{herwig2005}, the carbon is transported
from the core (spinstars) or from the surface (AGB stars) to the
H-burning shell where the CNO cycle is active, the carbon is transformed
into $^{13}$C and $^{14}$N. These signatures should be detectable in the
$^{12}$C/$^{13}$C isotopic ratio of a star. High $^{12}$C/$^{13}$C and
$\mathrm{[C/N]}$ ratios indicate only partial hydrogen burning by the
CNO cycle, while low $^{12}$C/$^{13}$C and $\mathrm{[C/N]}$ ratios are a
signature of more complete burning by the CNO cycle \citep{maeder2014}.

\cite{chiappini2008} calculated the predicted $^{12}$C/$^{13}$C isotopic ratio in the
primordial ISM from which the first low-mass stars formed, if the
first-generation stars were dominated by spinstars. They predict the
$^{12}$C/$^{13}$C isotopic ratio to be between 30-300, whereas if the first stars
were not dominated by spinstars, the ratio would be $\sim4500$ at
$\mathrm{[Fe/H]} = -3.5$ and as much as $\sim31000$ at $\mathrm{[Fe/H]}
= -5.0$. Models of mixing and fallback SNe events \citep{umeda2003,
nomoto2013}, suggested to occur in the early universe, also predicts
low $^{12}$C/$^{13}$C isotopic ratios, due to the mixing in the pre-supernova
evolution stage between the He convective shell and H-rich envelope
\cite{iwamoto2005}. However, \citet{maeder2014} predict differences in
the ratio due to the different physical conditions and timescales for
the production of $^{13}$C during the stellar evolution of spinstars or
in the SN explosion of the mixing and fallback models. 

\begin{figure}
\includegraphics[angle=0,width=3.5in]{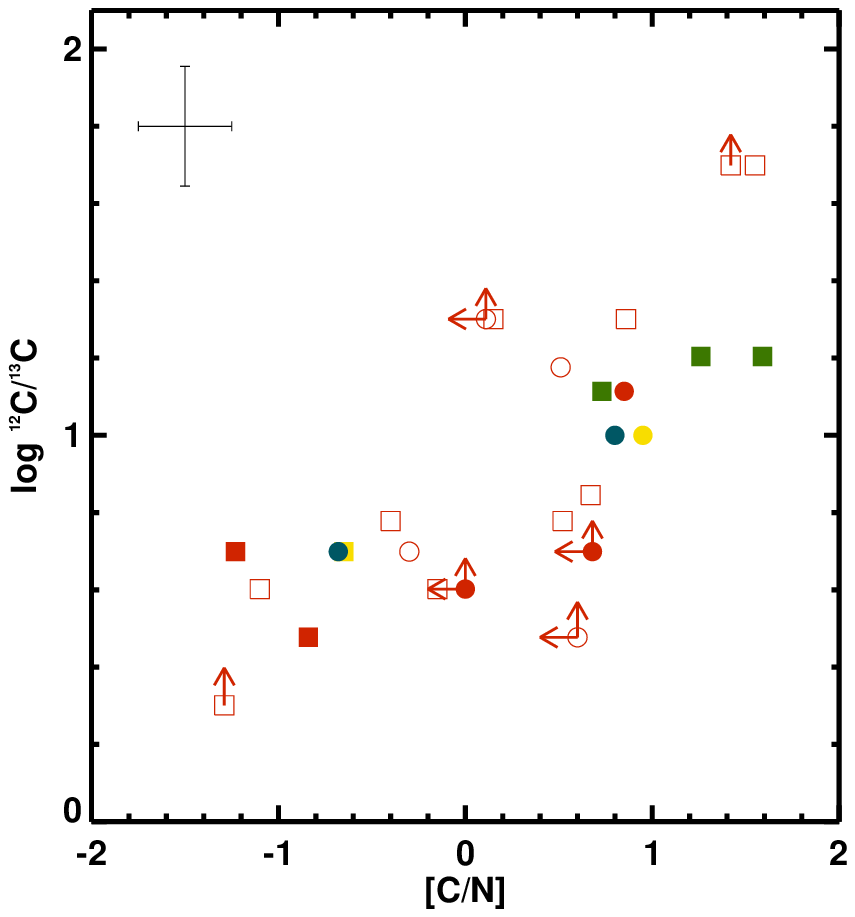}
\caption{$^{12}$C/$^{13}$C isotopic ratios, as a function of $\mathrm{[C/N]}$, for CEMP
stars in the combined sample and from \citet{norris2013b}. The squares
represent stars with $\log~g > 3.0$; dots represent stars with $\log~g <
3.0$. Filled symbols are from the combined sample; open symbols are from
\citet{norris2013b}. The color-coding of the symbols is as in Figure~\ref{fig3}. 
An approximate error bar for the sample stars is shown in the upper left
of the panel. 
\label{fig10}} 
\end{figure}

We have derived the $^{12}$C/$^{13}$C isotopic ratios for 11 of our stars, and
lower limits for an additional two stars. \cite{norris2013b} also
investigated $^{12}$C/$^{13}$C isotopic ratios in their sample of CEMP-$no$ stars
(which, except for BD+44$^{\circ}$493 and Segue~1-7, all belong to the
\citet{yong2013} sample), but they were able to derive isotopic ratios
for only 5 of their 15 stars; for the remaining they provided lower
limits. Figure~\ref{fig10} plots the $^{12}$C/$^{13}$C isotopic ratios, as a
function of $\mathrm{[C/N]}$ (uncorrected C abundances), for the CEMP stars in our sample and those
of \citet{norris2013b}. In this plot, dots represent stars with
$\log~g < 3.0$ in which some internal CNO cycle processing might have
changed the initial $^{12}$C/$^{13}$C isotopic ratios, and squares represent stars
with $\log~g > 3.0$ stars, which should have preserved their initial
$^{12}$C/$^{13}$C isotopic ratios. 

We find low ($\sim5$) $^{12}$C/$^{13}$C isotopic ratios for all of our CEMP-$no$
stars, consistent with the equilibrium value for CNO-cycle processed
material. This shows that the material from which these stars formed has
undergone mixing, whether in spinstars or in some pre-supernova
evolution. The $^{12}$C/$^{13}$C isotopic ratios found in the CEMP-$s$ stars of
our sample are generally higher ($\sim13$). This value is low enough to
be a signature of H-burning via the CNO cycle, which is also expected if
the carbon excess found in CEMP-$s$ stars are transferred from an AGB
companion, where multiple dredge up events mix the material in the star.
However, according to \citet{bisterzo2012}, current AGB models do not
include sufficient mixing to replicate the low $^{12}$C/$^{13}$C
isotopic ratios
found in CEMP-$s$ stars.

\subsection{Lithium}

Paper~I explored the lithium abundances detected in the UMP
CEMP-$no$ stars presented there, showing that all of these have Li
abundances below the plateau found for non carbon-enhanced metal-poor
dwarfs at $A(\rm Li) = 2.05$, the so-called Spite plateau
\citep{spite1982}. This result supplements a similar finding by
\citet{masseron2012}, and is consistent with the possible depletion of
Li by the progenitors of CEMP-$no$ stars suggested by \citet{piau2006}.
In this model, the first, presumably massive, stars that formed in the
universe are believed to have destroyed all of their Li. The observed Li
abundances of the EMP and UMP stars are expected to be the result of the
mixing of the Li-free material ejected from the first stars with the
unprocessed ISM having a Li abundance generated by Big Bang
Nucleosynthesis. In this sense the lithium abundances of metal-poor
main-sequence and subgiant stars can also be used to estimate the degree
to which the material from the source star has been diluted.\footnote{We
note that, although it is presented without attribution, the scenario
proposed by Bonifacio et al. (2015) to account for the ``meltdown'' of
the Spite Li plateau at low metallicities and the possible resolution of
the so-called cosmological Li problem (their Sec. 5.2) is essentially
the same as that proposed by Piau et al. (2006), as acknowledged by P.
Bonifacio (private comm. to TCB).} 
 
In this paper we present three additional CEMP-$no$ stars,
HE~0100$-$1622 with $A(\rm Li) < 1.12$, HE~0440$-$1049 with $A(\rm Li) =
2.00$, and HE~2331$-$7155 with $A(\rm Li) < 0.37$. The latter star is a
giant with $\log~g = 1.5$, so this star has most likely internally
depleted its initial lithium. The one star, HE~0440$-$1049, with a
higher Li abundance, close to the Spite plateau value also has the
lowest carbon-abundance ratio among these stars ($\mathrm{[C/Fe]} =
+0.69$).

\begin{figure}
\includegraphics[angle=0,width=3.5in]{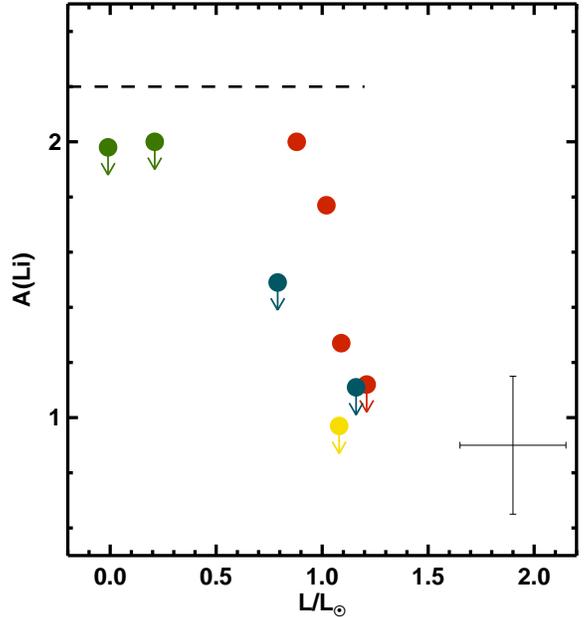}
\caption{Absolute Li abundances, $A(Li)$ and upper limits, indicated by
arrows, as a function of luminosity, for the dwarf and subgiant CEMP
stars in our sample. The Spite plateau lithium abundance ($A(\rm Li) = 2.2$)
is indicated with a dashed line. The color-coding of the symbols is as in
Figure~\ref{fig3}. An approximate error bar for the sample stars is
shown in the lower right of the panel. 
\label{fig11}} 
\end{figure}

We have also examined the Li abundances for the CEMP-$s$, CEMP-$r$ and
CEMP-$r/s$ stars in our new sample, but only upper limits could be
derived for these. Figure~\ref{fig11} shows the Li abundances and upper
limits detected for all the dwarf and subgiants in our sample, including
those presented in Paper~I, as function of luminosity
(\citealt{yong2013} did not present Li abundances for their stars). The
additional CEMP-$no$ stars follow the result from Paper~I and
\citet{masseron2012} that all CEMP-$no$ stars exhibit some level of Li
depletion with respect to the Spite plateau.

\subsection{Strontium and Barium}

The strontium and barium abundances for VMP stars have received a great
deal of attention over the past few years, in part because these two
species are often the only neutron-capture elements for which abundances
can be measured in the most metal-poor stars, making these two elements
our only clue to the nature of neutron-capture processes at the earliest
times in our Galaxy \citep{aoki2013a,hansen2013,roederer2014a}. 

\begin{figure}
\includegraphics[angle=0,width=3.5in]{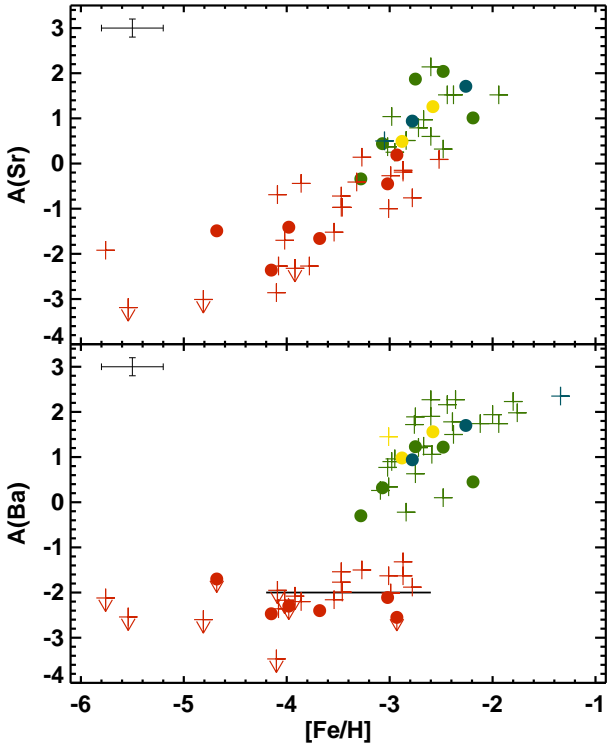}
\caption{Absolute abundances of Sr and Ba, as a function of $\mathrm{[Fe/H]}$, for
the CEMP stars. The suggested $A(\rm Ba)$ floor is indicated by the solid line. The
symbols and color-coding are as in Figure~\ref{fig9}. An approximate error bar
for the sample stars is shown in the upper left of each panel. 
\label{fig12}} 
\end{figure}

\begin{figure}
\includegraphics[angle=0,width=3.5in]{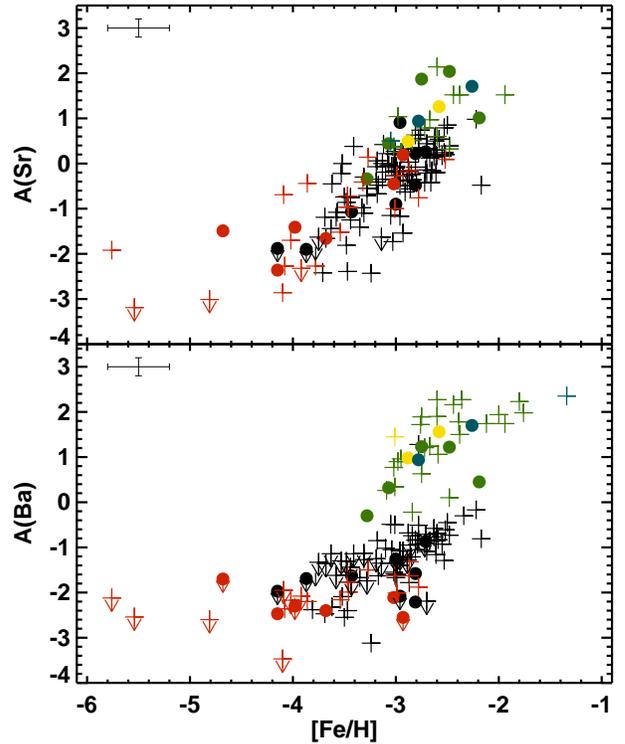}
\caption{Absolute abundances of Sr and Ba, as function of $\mathrm{[Fe/H]}$, for
all of the stars in our combined sample. The symbols and color-coding are as in
Figure~\ref{fig9}. An approximate error bar for the sample stars is shown in the upper left
of each panel.\label{fig13}} 
\end{figure}

We have obtained detections or strong upper limits for Sr and Ba for all
the stars in our sample, listed in Table~\ref{tab4}, Table~\ref{tab5},
Table~\ref{tab6} and Table~\ref{tab7}. Figure~\ref{fig12} and
Figure~\ref{fig13} shows the absolute Sr and Ba abundances as a function
of $\mathrm{[Fe/H]}$. Figure~\ref{fig12} shows only the CEMP stars, while
Figure~\ref{fig13} shows all of the stars in our combined sample.

Inspection of Figure~\ref{fig12} and Figure~\ref{fig13} indicates a
clear grouping of the different classes of stars considered in our
study. Recall that $\mathrm{[Ba/Fe]}$ is used to differentiate the
CEMP-$no$ stars from the CEMP-$s$ and CEMP-$r/s$ stars. The NMP (non
carbon-enhanced stars) exhibit a wide range of Ba abundances, from
$A(\rm Ba) \sim -4.0$ to $A(\rm Ba) \sim 0.0$, while all the CEMP-$no$
stars for which we have Ba detections exhibit Ba abundances of
$A(\rm Ba) \sim -2.0$, independent of metallicity (most clearly seen
when plotting only the CEMP stars, Figure~\ref{fig12}). In contrast, the
behavior of the Sr abundances for stars in our sample is substantially
different. The individual classes of the stars in our sample are mixed
together in a band showing decreasing $A(\rm Sr)$ with decreasing
$\mathrm{[Fe/H]}$, but with a possible change in the trend at the
lowest metallicity (around $\mathrm{[Fe/H]} \sim -4.2$). We emphasize
that the area below $\mathrm{[Fe/H] = -4}$ is only sparsely populated,
with most stars only having an upper limit on Sr and Ba. The current
data certainly suggest the presence of a floor in Ba at extremely low
metallicity, but not for Sr; more detections of both species are
strongly desired.

In the universe today, Ba is primarily produced by the main $s$-process
in lower-mass AGB stars \citep{busso1999}, but as this process was not
operating in the early universe, the Ba found in the CEMP-$no$ stars
must have some alternative origin. It was shown by
\cite{frischknecht2010} that spinstars can produce some amount of slow
neutron-capture elements such as Sr and Ba via the ``weak'' $s$-process,
provided some Fe seeds are available. It is possible that these elements
can also be produced by the mixing and fallback models, by ejection of a
tiny fraction of the heavy elements created in the explosion
\citep{takahashi2014}. Recently, \cite{roederer2014a} found four
CEMP-$no$ stars with clear $r$-process-element abundance patterns, confirming
the early onset of the rapid neutron-capture process in the Galaxy.
These authors also stress the need for Fe seeds for the weak $s$-process
to operate efficiently in spinstars, meaning it will not occur in a
completely metal-free star.

\subsection{Abundance Profiles}

The peculiar abundance patterns of the CEMP-$s$ stars, showing large
enhancements in carbon, nitrogen, and slow neutron-capture elements, are
believed to be the result of mass transfer from an AGB companion in a
binary system with the presently-observed low-mass metal-poor star.
Indeed, radial-velocity monitoring of CEMP-$s$ stars are consistent with
essentially all of these stars belonging to binary systems
\citep{lucatello2005}. Thus, the abundances observed in CEMP-$s$ stars
offer us a unique opportunity to constrain the properties of very
metal-poor AGB stars.

\begin{figure*}
\begin{center}
\includegraphics[width=6.5in]{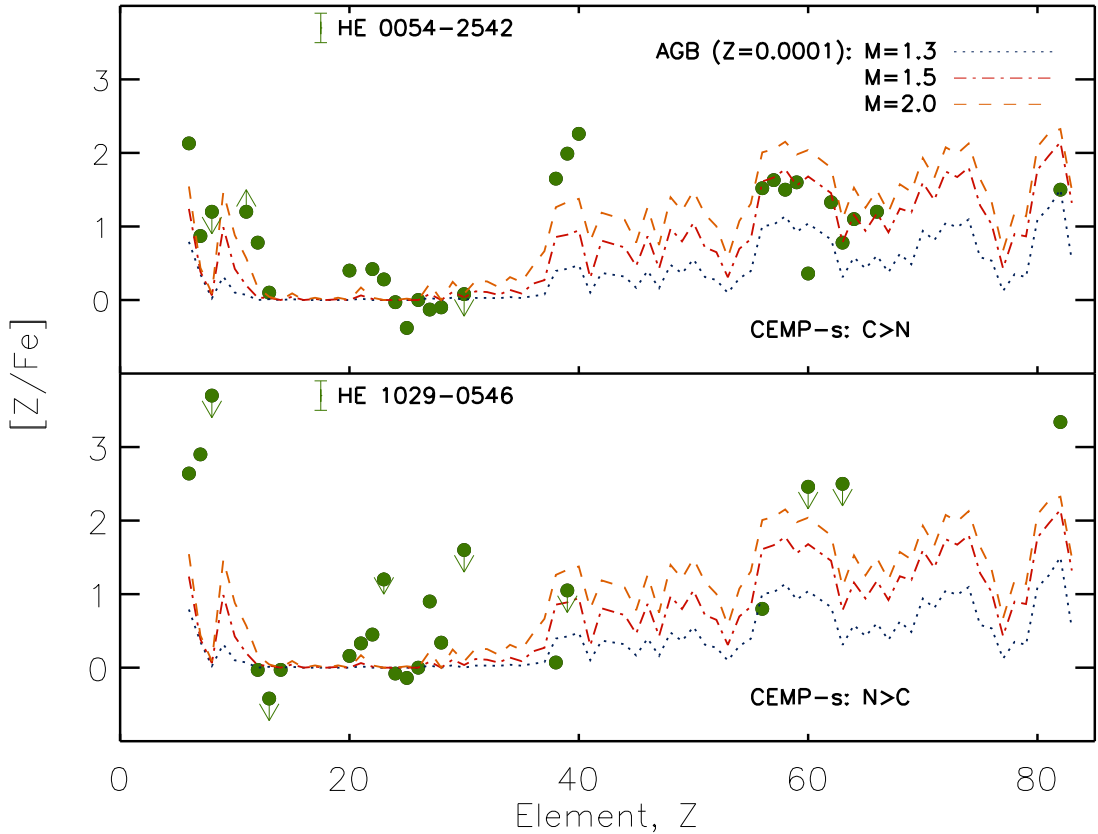}
\end{center}
\caption{Observed elemental-abundance patterns for two CEMP-$s$ stars, along with
predicted yields for metal-poor AGB models of three different masses,
1.3M$_{\odot}$, 1.5M$_{\odot}$, and 2.0M$_{\odot}$. A representative
error bar on the derived abundances is shown next to the star name in 
each panel. \label{fig14}} 
\end{figure*}

In the above division of stars into those with either $\mathrm{[C/N]}>0$
or $\mathrm{[C/N]}<0$, we see that the great majority of the CEMP-$s$
stars have $\mathrm{[C/N]}>0$. To further investigate the properties of
the AGB stars that created the elemental over-abundances detected in
CEMP-$s$ stars, Figure~\ref{fig14} shows the observed
elemental-abundance patterns of two CEMP-$s$ stars -- one having
$\mathrm{[C/N]}>0$ and one with $\mathrm{[C/N]}<0$ -- along with the
predicted yields from metal-poor ($Z=0.0001$) AGB models of three
different masses (1.3M$_{\odot}$, 1.5M$_{\odot}$, and 2.0M$_{\odot}$),
taken from the F.R.U.I.T.Y database \citep{cristallo2011,cristallo2009}. 

None of the models reproduce the large amounts of carbon and nitrogen
detected in these stars, and none of the models have $\mathrm{[C/N]}<0$.
The heavy neutron-capture elements for the CEMP-$s$ star with
$\mathrm{[C/N]}>0$ (\object{HE 0054-2542}) are well-fit by the M=1.5M$_{\odot}$ model, but none
of the models produce sufficient amounts of the light neutron-capture
elements (Sr,Y, and Zr) to match this star. The star with
$\mathrm{[C/N]}<0$ (\object{HE 1029-0546}) is also not well-fit by any of the models, but it
does exhibit a general lower enhancement in $s$-process elements than
the $\mathrm{[C/N]}>0$ star, pointing toward a lower mass AGB star as the
progenitor of \object{HE 1029-0546}.

\begin{figure*}
\includegraphics[angle=0,width=6.8in]{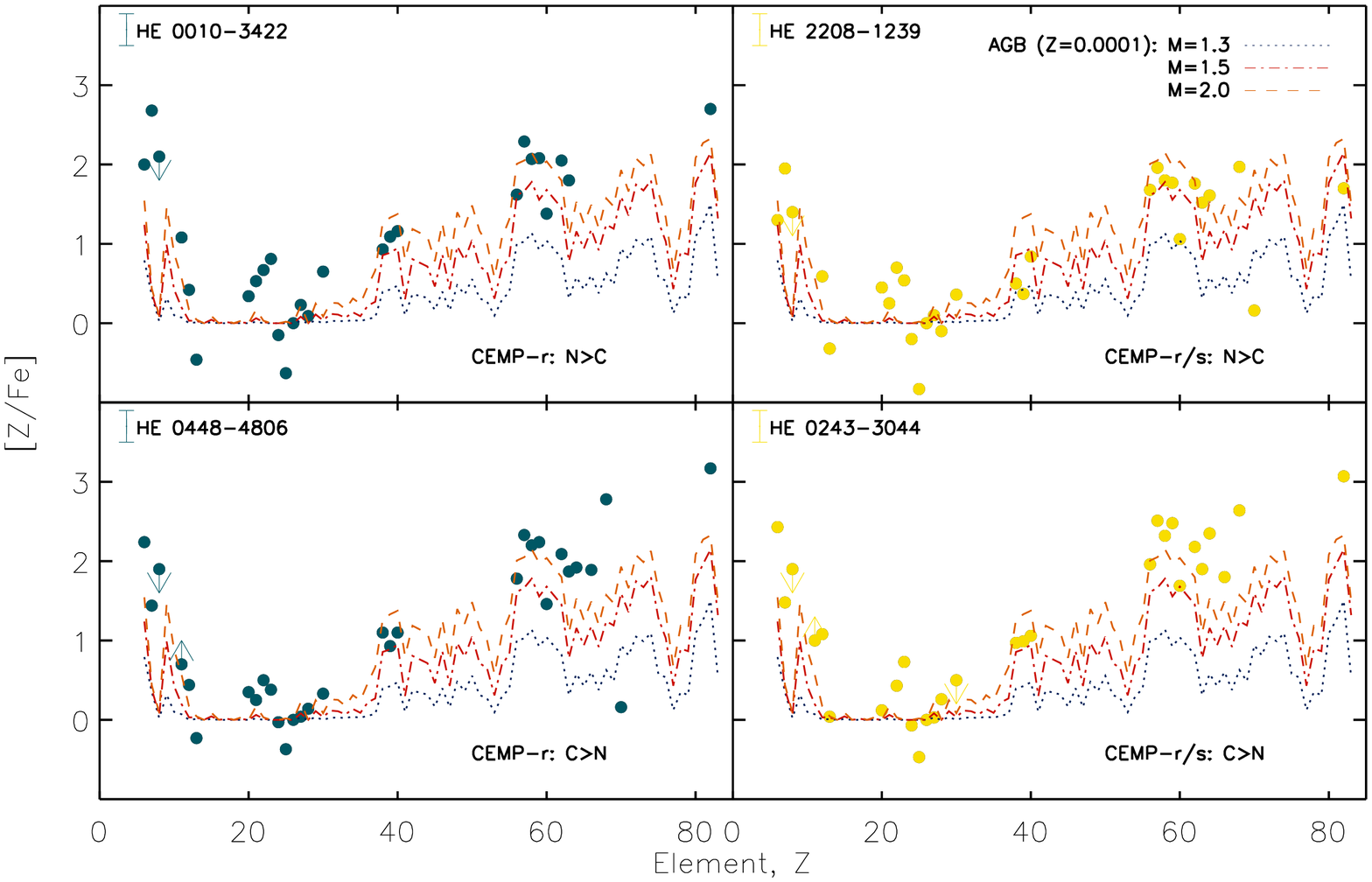}
\caption{Observed elemental-abundance patterns for the two CEMP-$r$
stars (left panels) and the two
CEMP-$r/s$ stars (right panels) in our sample, along with predicted yields for
metal-poor AGB stars of three different masses, 1.3M$_{\odot}$, 1.5M$_{\odot}$, and 2.0M$_{\odot}$.
A representative error bar on the derived abundances is shown next to
the star name in each panel.
\label{fig15}}
\end{figure*}

Figure~\ref{fig15} shows the observed elemental-abundance patterns of
the two CEMP-$r$ and CEMP-$r/s$ stars in our sample along with yields
from the same metal-poor AGB stars. For the CEMP-$r$ stars (left panels)
none of the AGB models reproduce the observed C and N abundances, and
while the most massive AGB model does reproduce some of the observed
abundances for the n-capture elements in HE~0010$-$3422, none of the
models are a good fit for the abundances observed in HE~0448$-$4806.
Considering the CEMP-$r/s$ stars (right panels), the two most massive
AGB models fit the observed C abundance for HE~2208$-$1239 and also some
of the n-capture element abundances observed in this star. For the other
CEMP-$r/s$ star, only the abundances for Sr, Y, and Zr can be reproduced
by an AGB model. 

It should also be noted that the material transferred from the AGB
star and onto the presently-observed CEMP star is expected to be mixed
with the original stellar material via thermohaline mixing
\citep{stancliffe2007, stancliffe2008}. The amount of dilution for the
transferred material is not presently well-constrained, hence we have
not included it in the comparison between observed abundances and AGB
yields.

\section{Summary and Discussion}

Below we summarize and discuss the abundance signatures found for the
CEMP stars in our sample, and what can be learned about the different
progenitors that are currently suggested to account for the known
sub-classes of CEMP stars. However, one result which is consistently
found in all abundance analysis of very metal-poor stars, both
carbon-enhanced and non-carbon-enhanced \citep[see][]{cayrel2004,
yong2013}, and which we confirm with our sample, is the small
over-abundance detected for the $\alpha$-elements ($<\mathrm{[Ca,
Ti/Fe]}> \sim +0.35$) and the very low star-to-star scatter for both the
$\alpha$-elements ($\sim 0.15$ dex) and the iron-peak elements ($\sim
0.20$ dex).

\subsection{CEMP-$no$ Stars}

The progenitors for the CEMP-no stars are thought to be either the massive
fast-rotating stars \citep[spinstars;][]{meynet2006,hirschi2007,maeder2014},
the proposed mixing and fallback supernovae \citep{umeda2003,nomoto2013}, or possibly contributions from both. Both
models explain well the observed large over-abundances of carbon and
nitrogen. In the spinstar scenario, a star that exhibits $\mathrm{[C/N]}
> 0$ is the result of incomplete hydrogen burning via the CNO cycle,
followed by mild mixing, whereas a star with $\mathrm{[C/N]} < 0$ is the
sign of more complete hydrogen burning \citep{maeder2014}. Other
abundance signatures detected in some CEMP-$no$ stars, such as low
$^{12}$C/$^{13}$C isotopic ratios, which we find in all the CEMP-$no$ stars where
we could measure this ratio, and high Na, Mg and Al abundances, as
detected in two of the CEMP-$no$ stars with $\mathrm{[C/N]} < 0$, can
also be explained, according to \cite{maeder2014}, as the result of more
mixing and processing of material in the spinstar. 

\cite{tominaga2014} have used ``profile fitting'' to show that
the yields from mixing and fallback SNe also well-fit the observed
abundance patterns of CEMP-$no$ stars. With a range of explosion
energies and mass cuts, and by including mixing in some of the the
models, the authors fit the abundance profiles of 12 CEMP-$no$ stars.
\citet{tominaga2014} also point out that the mixing and fallback SN
model fits all the observed elements up to mass number $Z = 30$,
including the $\alpha$-elements and the iron-peak elements, while the
spinstar models require a complementary SN contribution to create these
elements in the appropiate proportion.

In our combined sample of 30 CEMP-$no$ stars, we observe stars with both
$\mathrm{[C/N]} > 0$ and $\mathrm{[C/N]} < 0$, but stars with
$\mathrm{[C/N]} < 0$ are only found at extremely low metallicity,
$\mathrm{[Fe/H]} < -3.4$, a metallicity below which the CEMP-$no$ stars
with large enhancements in Na, Mg and Al also are found; above this
metallicity the abundance spread is smaller. This indicates that the
large degrees of internal mixing and processing required to produce the abundance
pattern seen in stars such as HE~1327$-$2326 \citep{frebel2006} and
HE~2323$-$0256 \citep{yong2013} was only operating at the very earliest times. 

Our data also supports the claim of \citet{spite2013} and \citet{bonifacio2015} for the presence
of two carbon ``bands'' that comprise the distribution of the absolute
carbon abundances for CEMP stars, although with a smoother transition
between the bands than was found by these authors. The majority of the
CEMP-$no$ stars have carbon abundances falling on the lower band ($A(\rm
C)\sim6.5$), but a few with metalicities above $\mathrm{[Fe/H]} > -3.0$
have carbon abundances in the higher band ($A(\rm C) \sim 8.5$),
indicating a different origin for the carbon found in these three stars
as opposed to those with carbon abundances in the lower band. 

We have inspected the $^{12}$C/$^{13}$C isotope ratios for the CEMP stars in our
sample, and find low $^{12}$C/$^{13}$C ratios for the CEMP-$no$ stars we
observed; again a signature of mixing in the progenitor stars. As
noted above, both the spinstar models and the mixing and fallback models
produce small $^{12}$C/$^{13}$C ratios, but so far no values for
expected $^{12}$C/$^{13}$C ratios has been published for the SNe models,
making a comparison difficult. 

Finally, we detect a ``floor'' in the absolute Ba abundances of CEMP-$no$
stars, at an abundance level $A(\rm Ba)\sim-2$. This plateau only exists for
the CEMP-$no$ stars, and must as such be an unique signature for the
progenitors of these stars. As mentioned above, both the spinstar and
the mixing and fallback models can produce some amount of
neutron-capture elements. For the mixing and fallback models no yields have
yet been published. In any case, the spinstars require Fe seeds, hence
some prior SNe pollution is needed.    

\citet{roederer2013} asked the question if {\it any} stars, no
matter at what low metallicity, completely lack neutron-capture
elements. The CEMP-$no$ stars are thought to be among the first low-mass
stars to have formed, and although not all have detected neutron-capture
elements, the discovery of a floor for the absolute Ba abundance of
CEMP-$no$ stars at extremely low metallicity supports the interpretation
that some mechanism producing neutron-capture elements was present very
early in the Galaxy.

\cite{norris2013b} suggested that, in order to distinguish between spinstars
and the mixing and fallback SNe for CEMP-$no$ stars, one needs to
investigate the abundances in these stars for elements that are produced
in the deeper layers of the progenitor stars, such as Si and Ca. For our
sample, the Ca abundances follow the general $\alpha$-element trends of
VMP halo stars, and only a few of our stars have detections for Si. More
predicted elemental yields, over the full range of elements -- both
light and heavy, along with predicted $^{12}$C/$^{13}$C ratios for
both sets of progenitor models, and additional CEMP-$no$ stars with
measured Si abundances, are required to resolve this issue.

\subsection{CEMP-$s$ Stars}

The origin of the CEMP-$s$ stars has for some time been ascribed to
mass transfer in a binary system from a now extinct AGB star, however
there are still abundance signatures found in CEMP-$s$ stars which the
AGB models have difficulties in explaining. 

For the 32 CEMP-$s$ stars in our combined sample, we find that the
great majority exhibit larger C-enhancements than N-enhancements, i.e.,
$\mathrm{[C/N]}>0$. However, we do find a few CEMP-$s$ stars with
$\mathrm{[C/N]}<0$, suggesting a higher degree of H-burning via the CNO
cycle has occurred in the AGB companion in such cases. We also find
higher $^{12}$C/$^{13}$C isotopic ratios for the CEMP-$s$ stars in this sample
than found for the CEMP-$no$ stars. However, the observed ratios are
sufficiently low to require extensive mixing, which is also expected
in an AGB star. Nevertheless, the observed $^{12}$C/$^{13}$C ratios for
CEMP-$s$ stars are not reproduced by the AGB models \citep{bisterzo2012}.

The CEMP-$s$ stars in our sample have carbon abundances that place them
on the higher of the two carbon bands of \citet{spite2013}, shown in
Figure~\ref{fig9}. Thus, there appears to be a maximum carbon abundance
attained for the CEMP-$s$ stars, which provides a constraint on the
efficiency of the mass transfer and/or the production of C in AGB stars
at very low metallicity.

\subsection{CEMP-$r$ and CEMP-$r/s$ Stars}

Small numbers of CEMP-$r/s$, and even fewer CEMP-$r$ stars, are known,
but we have identified two of each class in our sample. The first
CEMP-$r$ star to be found was CS~22892-052 \citep{mcwilliam1995,sneden2003}. This star
is both classified as an $r$-II star\footnote{$\mathrm{[Eu/Fe]} > +1.0$
and $\mathrm{[Ba/Eu]} < 0.0$ \citep{beerschristlieb2005}}, and a
CEMP-$r$ star, and has an abundance pattern for the heavy $r$-process
elements that well-fits the scaled solar system $r$-process abundance pattern.
It is not known with certainty what astrophysical site produces the
carbon over-abundances for CEMP-$r$ stars, but radial-velocity
monitoring of CS~22892-052 shows that this star is unlikely to be in a
binary system \citep{hansen2011}, suggesting that the carbon
enhancements seen for CEMP-$r$ stars are not the result of mass transfer
in a binary system, but, more likely (as in the case of the CEMP-$no$
stars), that they were born from an ISM that was previously polluted with
carbon.

For CEMP-$r/s$ stars, which exhibit contributions from both the $r$- and
$s$-process, it has been proposed that they were born with their
$r$-process-element abundances, and then gain their carbon and
$s$-process-element abundances via mass transfer in a binary system
\citep{qianwasserburg2003}. Recently, it has also been suggested that
these stars could be the result of the $i$-process, a process that is
intermediate between the $r$- and the $s$-process, and thought to occur
in high-mass ``super-AGB'' stars \citep{bertolli2013}.

The general abundance patterns we detect in the two CEMP-$r$ and two
CEMP-$r/s$ stars in our sample follow those of the other CEMP stars.
However, we find both sub-classes present on both the high and the low
carbon-abundance bands shown in Figure~\ref{fig9}. We also find
$^{12}$C/$^{13}$C isotopic ratios in these stars that match both what is found
for the CEMP-$s$ stars and for the CEMP-$no$ stars. Larger samples of
both CEMP-$r$ and CEMP-$r/s$ stars with available high-resolution,
high-S/N spectra are clearly needed to obtain a more secure picture of
the likely progenitor(s) of these objects. We have shown that one of
the CEMP-$r$ stars and one of the CEMP-$r/s$ stars may be classified as
NEMP stars; confirmation of their binary-pollution origin awaits future
radial-velocity monitoring. 

\subsection{Outlook}

We require still larger samples of the variety of low-metallicity
stars presented in this paper to distinguish between the abundance
patterns of stars representing the general trends and those that are
just peculiar outliers. High-resolution spectroscopic follow-up of
several large surveys, such as SkyMapper \citep{keller2007,
  keller2014,jacobson2015}, TOPoS \citep{caffau2013}, based on stars selected
by SDSS/SEGUE \citep{yanny2009} \citep[see also][]{aoki2013b}, the CEMP-star
searches from the HK survey, the HES, and the RAVE survey
\citep{steinmetz2006} described by \citet[][and in
  preparation]{placco2010,placco2011,placco2013,placco2014a}, and stars
selected from LAMOST \citep{deng2012,li2015} have increased in recent years,
and need to be expanded further. These surveys will soon provide more examples
of CEMP-$r$ and CEMP-$r/s$ stars, for which our current samples are very
limited. They will also enlarge the numbers of known CEMP-$s$ and CEMP-$no$
sub-classes of carbon-enhanced stars. With detailed and homogeneous analyses
of these stars, we can look forward to detecting the elemental abundance
signatures that constrain the nature and sites of the nucleosynthesis events
that first enriched the Milky Way.

\acknowledgments

This work was supported by Sonderforschungsbereich SFB 881 ``The Milky
Way System'' (subproject A4) of the German Research Foundation (DFG).
T.C.B. and V.P.M acknowledges partial support for this work from grants PHY
08-22648; Physics Frontier Center/{}Joint Institute or Nuclear
Astrophysics (JINA), and PHY 14-30152; Physics Frontier Center/{}JINA
Center for the Evolution of the Elements (JINA-CEE), awarded by the US
National Science Foundation. C.J.H.was supported by a research grant
(VKR023371) from VILLUM FONDATION and by Sonderforschungsbereich SFB 881 "The
Milky Way System" (subproject A5) of the German Research Foundation
(DFG). A.F. is supported by NSF CAREER grant AST-1255160. M.A., M.S.B.,
J.E.N., and D.Y. acknowledge support from the Australian Research Council
(grants DP0342613, DP0663562 and FL110100012) for studies of the Galaxy's most
metal-poor stars. Furthermore, we thank the referee for helpful comments.

{\it Facilities:} \facility{VLT:Kueyen}, \facility{AAT}, \facility{ATT}, \facility{Blanco},
\facility{CTIO:1.5m}, \facility{ESO:3.6m}, \facility{Mayall}, \facility{SOAR},\facility{UKST}

\clearpage

\begin{thebibliography}{}

\bibitem[Allen et al.(2012)]{allen2012} Allen, D. M., Ryan, S. G., Rossi, S., Beers,
  T. C., \& Tsangarides, S. A., 2012, \aap, 548, A34

\bibitem[Allende Prieto et al.(2004)]{allende2004} Allende Prieto, C.,
  Barklem, P.S., Lambert, D.L. \& Cunha, K., 2004, \aap, 420, 183

\bibitem[Alonso et al.(1996)]{alonso1996} Alonso, A., Arribas, S., \&
  Mart{\'{\i}}nez-Roger, C., 1996, \aap, 313, 873

\bibitem[Alonso et al.(1999)]{alonso1999} Alonso, A., Arribas, S., \&
  Mart{\'{\i}}nez-Roger, C., 1999, \aaps, 140, 261


\bibitem[Aoki et al.(2006)]{aoki2006} Aoki, W., Frebel, A., Christlieb, N., et al.,
  2006, \apj, 639, 897

\bibitem[Aoki et al.(2013b)]{aoki2013b}	Aoki, W., Beers, T.C., Lee,
Y.S., et al. 2013, \aj, 145, 13

\bibitem[Aoki et al.(2002a)]{aoki2002a} Aoki, W., Norris, J. E., Ryan, S. G.,
  Beers, T. C., \& Ando, H., 2002, \apjl, 576, L141

\bibitem[Aoki et al.(2002b)]{aoki2002b} Aoki, W., Ryan, S. G., Norris, J. E., et al.,
  2002, \apj, 580, 1149

\bibitem[Aoki et al.(2013a)]{aoki2013a} Aoki, W., Suda, T., Boyd, R. N., Kajino,
  T., \& Famiano, M. A., 2013, \apjl, 766, L13 

\bibitem[Asplund et al.(2009)]{asplund2009} Asplund, M., Grevesse, N.,
  Sauval, A. J., \& Scott, P., 2009, \araa, 47, 481

\bibitem[Barbuy et al.(2005)]{barbuy2005} Barbuy, B., Spite, M., Spite, F.,
  2005, \aap, 429, 1031 

\bibitem[Beers \& Christlieb(2005)]{beerschristlieb2005} Beers, T. C., \&
  Christlieb, N., 2005, \araa, 43, 531

\bibitem[Beers et al.(2007)]{beers2007} Beers, T. C., Flynn, C., Rossi, S., et
  al., 2007, \apjs, 168, 128 

\bibitem[Beers et al.(1992)]{beers1992} Beers, T. C., Preston, G. W., \&
  Shectman, S. A., 1992, \aj, 103, 1987

\bibitem[Bertolli et al.(2013)]{bertolli2013} Bertolli, M. G., Herwig, F.,
  Pignatari, M., \& Kawano, T., 2013, arXiv:1310.4578

\bibitem[Bessell(2005)]{bessell2005} Bessell, M. S., 2005, \araa, 43, 293

\bibitem[Bessell(2007)]{bessell2007} Bessell, M. S., 2007, \pasp, 119, 605

\bibitem[Bisterzo et al.(2012)]{bisterzo2012} Bisterzo, S., Gallino, R.,
  Straniero, O., Cristallo, S., \& K{\"a}ppeler, F., 2012, \mnras, 422, 849 

\bibitem[Bonifacio et al.(2015)]{bonifacio2015} Bonifacio, P., Caffau, E.,
  Spite, M., et al., 2015, arXiv:1504.05963

\bibitem[Bonifacio et al.(2000)]{bonifacio2000} Bonifacio, P., Monai, S., \&
  Beers, T. C., 2000, \aj, 120, 2065

\bibitem[Bromm \& Loeb(2003)]{bromm2003} Bromm, V., \& Loeb, A., 2003, \nat,
  425, 812

\bibitem[Busso et al.(1999)]{busso1999} Busso, M., Gallino, R., \& Wasserburg,
  G. J., 1999, \araa, 37, 239

\bibitem[Caffau et al.(2013)]{caffau2013} Caffau, E., Bonifacio, P., Sbordone,
  L., et al., 2013, \aap, 560, A71

\bibitem[Castelli \& Kurucz(2003)]{castelli2003} Castelli, F. \& Kurucz,
  R. L., 2003, IAU Symposium, 210

\bibitem[Cayrel et al.(2004)]{cayrel2004} Cayrel, R., Depagne, E., Spite, M.,
  et al., 2004, \aap, 416, 1117

\bibitem[Chiappini et al.(2008)]{chiappini2008} Chiappini, C., Ekstr{\"o}m,
  S., Meynet, G., et al., 2008, \aap, 479, L9

\bibitem[Christlieb et al.(2004)]{christlieb2004} Christlieb, N., Gustafsson,
  B., Korn, A. J., et al., 2004, \apj, 603, 708

\bibitem[Christlieb et al.(2008)]{christlieb2008} Christlieb, N., Sch{\"o}rck,
  T., Frebel, A., et al., 2008, \aap, 484, 721

\bibitem[Cohen et al.(2008)]{cohen2008} Cohen, J. G., Christlieb, N.,
  McWilliam, A., et al., 2008, \apjl, 672, 320

\bibitem[Cohen et al.(2013)]{cohen2013} Cohen, J. G., Christlieb, N.,
  Thompson, I., et al., 2013, \apj, 2013, 778, 56

\bibitem[Cohen et al.(2006)]{cohen2006} Cohen, J. G., McWilliam, A., Shectman,
  S., 2006, \aj, 132, 137

\bibitem[Cristallo et al.(2011)]{cristallo2011} Cristallo, S., Piersanti, L.,
  Starniero, O., et al., 2011, \apjs, 197, 17 

\bibitem[Cristallo et al.(2009)]{cristallo2009} Cristallo, S., Starniero, O.,
  Gallino, R., et al., 2009, \apj, 696, 797

\bibitem[Dekker et al.(2000)]{dekker2000} Dekker, H., D'Odorico, S., Kaufer,
  A., Delabra, B., \& Kotzlowski, H., 2000, Optical and IR Telescope
  Instrumentation and Detectors, 4008, 534

\bibitem[Demarque et al.(2004)]{demarque2004} Demarque, P., Woo, J. H., Kim,
  Y. C., \& Yi, S. K., 2004, \apjs, 155, 667 

\bibitem[Deng et al.(2012)]{deng2012} Deng, L., Newberg, H. J., Liu, C., et al. 2012, RAA, 12, 735

\bibitem[Dopita et al(2007)]{dopita2007} Dopita, M., Hart, J., McGregor, P.,
  et al., 2007, \apss, 310, 255

\bibitem[Eggleton (1971)]{eggleton1971} Eggleton, P.P., 1971, \mnras, 151, 351

\bibitem[Fran{\c c}ois et al.(2007)]{francois2007} Fran{\c c}ois , P.,
  Depagne, E., Hill, V., et al., 2007, \aap, 476, 935

\bibitem[Frebel et al.(2006)]{frebel2006} Frebel, A., Christlieb, N., Norris,
  J.E., et al., 2006, \apj, 652, 1585

\bibitem[Frebel et al.(2007)]{frebel2007} Frebel, A., Johnson, J. L., \&
  Bromm, V., 2007, \mnras, 380, L40

\bibitem[Frebel \& Norris (2015)]{frebelnorris2015} Frebel, A., \& Norris,
  J. N., 2015, arXiv:1501.06921

\bibitem[Frischknecht et al.(2010)]{frischknecht2010} Frischknecht U.,
  Hirschi, R., Meynet, G., et al., 2010, \aap, 522, A39

\bibitem[Gustafsson et al.(2008)]{gustafsson2008} Gustafsson, B., Edvardsson,
  B., Eriksson, N., et al., 2008, \aap, 486, 951

\bibitem[Hansen et al.(2013)]{hansen2013} Hansen, C. J., Bergemann, M.,
  Cescutti, G., et al., 2013, \aap, 551, A57

\bibitem[Hansen et al.(2011)]{hansen2011} Hansen T., Andersen, J.,
  Nordstr{\"o}m, B., Buchhave, L. A., \& Beers, T. C., 2011, \apjl, 743, L1

\bibitem[Hansen et al.(2014)]{hansen2014} Hansen, T., Hansen, C. J.,
  Christlieb, N., et al., 2014, \apj, 787, 162

\bibitem[Herwig(2005)]{herwig2005} Herwig, F., 2005, \araa, 43, 435

\bibitem[Hirschi(2007)]{hirschi2007} Hirschi, R., 2007, \aap, 461, 571

\bibitem[H{\o}g et al.(2000)]{hoeg2000} H{\o}g, E., Frabricius, C., Makarov,
  V.V., et al., 2000, \aap, 355, L27

\bibitem[Iwamoto et al.(2005)]{iwamoto2005} Iwamoto, N., Umeda, H., Tominaga,
  N., Nomoto, K., \& Maeda, K., 2005, Science, 309, 451

\bibitem[Jacobson et al.(2015)]{jacobson2015} Jacobson, H., Frebel, A.,
et al., \apj, in press

\bibitem[Johnson et al.(2007)]{johnson2007} Johnson, J. A., Herwig, F.,
  Beers., \& Christlieb, N., 2007, \apj, 658, 1203

\bibitem[Keller et al.(2014)]{keller2014} Keller, S. C., Bessel, M. S., Frebel,
  A., et al., 2014, \nat, 

\bibitem[Keller et al.(2007)]{keller2007} Keller, S. C., Schmidt, B.P., Bessel,
  M.S., et al., 2007, PASA, 24, 1 

\bibitem[Kupka et al.(2000)]{kupka2000} Kupka, F.G., Ryabchikova, T. A.,
  Piskunov, N. E., Stempels, H. C., \& Weiss, W. W., 2000, Baltic Astronomy, 9, 590

\bibitem[Kurucz(1995)]{kurucz1995} Kurucz R. L., 1995, Astrophysical
  Applications of Powerful New Databases, 78

\bibitem[Li et al.(2015)]{li2015} Li, H., Zhao, G., Christlieb, N., et al.,
  2015, \apj, 798, 110

\bibitem[Lucatello et al.(2005)]{lucatello2005} Lucatello, S., Tsangarides,
  S., Beers, T. C., et al., 2005, \apj, 625, 825

\bibitem[Maeder et al.(2014)]{maeder2014} Maeder, A., Meynet, G., \&
  Chiappini, C., 2014, arXiv:1412.5754

\bibitem[Masseron(2006)]{masseron2006} Masseron, T., 2006, P.h.D Thesis,
  Observatoire de Paris

\bibitem[Masseron et al.(2012)]{masseron2012} Masseron, T., Johnson, J. A.,
  Lucatello, S., et al., 2012, \apj, 751, 14

\bibitem[Masseron et al.(2014)]{masseron2014} Masseron, T., Plez, B., Van Eck,
  S., et al., 2014, \aap, 571, A47

\bibitem[McWilliam et al.(1995)]{mcwilliam1995} McWilliam, A., Preston, G. W.,
  Sneden, C., \& Searle, L., 1995, \aj, 109, 2757

\bibitem[Meynet et al.(2006)]{meynet2006} Meynet, G., Ekstr{\"o}m,
  S., \& Maeder, A., 2006, \aap, 447, 623

\bibitem[Nomoto et al.(2013)]{nomoto2013} Nomoto, K., Kobayashi, C., \&  Tominaga, N., 2013, \araa, 51, 457

\bibitem[Norris et al.(2013a)]{norris2013a} Norris, J. E., Bessell, M. S., Yong,
  D., et al., 2013, \apj, 762, 25

\bibitem[Norris et al.(2007)]{norris2007} Norris, J. E., Christlieb, N., Korn,
  A. J., et al., 2007, \apj, 670, 774

\bibitem[Norris et al.(1996)]{norris1996} Norris, J. E., Ryan, S. G., \&
  Beers, T. C., 1996, \apjs, 107, 391

\bibitem[Norris et al.(2013b)]{norris2013b} Norris, J. E., Yong, D., Bessell,
  M. S., et al., 2013, \apj, 762, 28

\bibitem[Piau et al.(2006)]{piau2006} Piau, L., Beers, T. C., Balsara, D. S., et
  al., 2006, \apj, 653, 300

\bibitem[Placco et al.(2013)]{placco2013} Placco, V. M., Frebel, A.,
  Beers, T. C., et al., 2013, \apj, 770, 104

\bibitem[Placco et al.(2014b)]{placco2014b} Placco, V.M., Frebel, A., Beers,
  T.C., \& Stancliffe, R.J., 2014, \apj, 797, 21 

\bibitem[Placco et al.(2014a)]{placco2014a} Placco, V.M., Frebel, A.,
Beers, T.C., et al. 2014, \apj, 781, 40 

\bibitem[Placco et al.(2011)]{placco2011} Placco, V.M., Kennedy, C.R., Beers,
  T.C., et al., 2011, \aj, 142, 188 

\bibitem[Placco et al.(2010)]{placco2010} Placco, V.M., Kennedy, C.R., Rossi,
  S., et al., 2010, \aj, 139, 1051 

\bibitem[Preston \& Sneden(2001)]{prestonsneden2001} Preston, G. W., \& Sneden,
  C., 2001, \aj, 122, 1545

\bibitem[Qian \& Wasserburg(2003)]{qianwasserburg2003} Qian, Y. Z., \& Wasserburg,
  G. J., 2003, \apj, 588, 1099

\bibitem[Reddy et al.(2003)]{reddy2003} Reddy, B. E., Tomkin, J., Lambert,
  D. L., \& Allende Prieto, C., 2003, \mnras, 340, 304

\bibitem[Roederer(2013)]{roederer2013} Roederer, I. U., 2013, \aj, 145, 26

\bibitem[Roederer et al.(2014a)]{roederer2014a} Roederer I. U., Preston,
  G. W., Thompson, I. B., Shectman, S. A., \& Sneden, C., 2014, \apj, 784, 158

\bibitem[Roederer et al.(2014b)]{roederer2014b} Roederer I. U., Preston,
  G. W., Thompson, et al., 2014, \aj, 147, 136

\bibitem[Rossi et al.(2005)]{rossi2005} Rossi, S., Beers, T.C., Sneden, C., et
  al., 2005, \apj, 130, 2804

\bibitem[Schlegel et al.(1998)]{schlegel1998} Schlegel, D. J., Finkbeiner,
  D. P., \& Davis, M., 1998, \apj, 500, 525

\bibitem[Simons et al.(1989)]{simons1989} Simons, J.W., Palmer, B.A., Hof,
  D.E., \& Oldenborg, R.C., 1989, Journal of the Optical Society of America B
  Optical Physics, 6, 1097 

\bibitem[Sneden(1973)]{sneden1973} Sneden, C., 1973, \apj, 184, 839

\bibitem[Sneden et al.(2003)]{sneden2003} Sneden, C., Cowan, J. J., Lawler,
  J. E., et al., 2003, \apj, 591, 936

\bibitem[Sobeck et al.(2011)]{sobeck2011} Sobeck, J. S., Kraft, R. P., Sneden,
  C., et al., 2011, \aj, 141, 175

\bibitem[Skrutskie et al.(2006)]{skrutskie2006} Skrutskie, M.F., Cutri, R.M.,
  Stiening, R., et al., 2006, \aj, 131, 1163

\bibitem[Spite et al.(2013)]{spite2013} Spite, M., Caffau, E., Bonifacio,
  P., et al., 2013, \aap, 552, A107

\bibitem[Spite \& Spite(1982)]{spite1982} Spite, F., \& Spite, M., 1982, \aap,
  115, 357

\bibitem[Stancliffe et al.(2009)]{stancliffe2009} Stancliffe, R.J., Church,
  R.P., Angelou, G.C., \& Lattanzio, J.C., 2009, \mnras, 396, 2313 

\bibitem[Stancliffe \& Eldridge (2009)]{stancliffeeldridge2009} Stancliffe
  R.J., \& Eldridge, J.J., 2009, \mnras, 396, 1699

\bibitem[Stancliffe \& Glebbeek (2008)]{stancliffe2008} Stancliffe, R., \&
  Glebbeek, E., 2008, \mnras, 389, 1828

\bibitem[Stancliffe et al.(2007)]{stancliffe2007} Stancliffe, R., Glebbeek,
  E., Izzard, R.G., \& Pols., O.R., 2007, \aap, 464, L57

\bibitem[Steinmetz et al.(2006)]{steinmetz2006}	Steinmetz, M., Zwitter,
T., Siebert, A., et al. 2006, \aj, 132, 1645 

\bibitem[Takahashi et al.(2014)]{takahashi2014} Takahashi, K., Umeda, H., \&
  Yoshida, T., 2014, \apj, 794, 40

\bibitem[Tominaga et al.(2014)]{tominaga2014} Tominaga, N., Iwamoto, N., \&
  Nomoto, K., 2014, \apj, 785, 98

\bibitem[Umeda \& Nomoto(2003)]{umeda2003} Umeda, H., \& Nomoto, K., 2003,
  \nat, 422, 871

\bibitem[Yanny et al.(2009)]{yanny2009} Yanny, B., Newberg, H.-J., Johnson, J. A., et al. 2009,
\aj, 137, 4377

\bibitem[Yong et al.(2013)]{yong2013} Yong, D., Norris, J. E., Bessell, M. S., et
  al., 2013, \apj, 762, 26

\end{thebibliography}
\end{document}